\documentclass[12pt]{article}
\usepackage{amsfonts,amssymb,amsmath,mathrsfs,latexsym,bbm,dsfont,amsthm}
\usepackage{color}
\usepackage{graphicx}
\usepackage[scriptsize,nooneline,hang]{caption}
\usepackage[hang,nooneline,scriptsize]{subfigure}
\usepackage{rotating}
\usepackage[all]{xy}
\begin{document}

\begin{center}
{\large \bf  Charged Dilation Black Holes as Particle Accelerators}
\end{center}

\vskip 5mm

\begin{center}
{\Large{Parthapratim Pradhan\footnote{E-mail: pppradhan77@gmail.com}}}
\end{center}

\vskip  0.5 cm

{\centerline{\it Department of Physics}}
{\centerline{\it Vivekananda Satabarshiki Mahavidyalaya}}
{\centerline{\it Manikpara, West Midnapur}}
{\centerline{\it West Bengal~721513, India}}

\vskip 1cm

\begin{abstract}
We examine the possibility of arbitrarily high energy in the Center-of-mass(CM) frame
of colliding neutral particles in the vicinity of the horizon of a charged dilation
black hole(BH). We show that it is possible to achieve the infinite energy in the background of the dilation black hole without fine-tuning of the angular momentum parameter. It is found that the CM energy $({\cal E}_{cm})$ of collisions of particles
near the infinite red-shift surface of the extreme dilation BHs are arbitrarily large while the non-extreme charged dilation BHs have the finite energy. We have also compared the ${\cal E}_{cm}$  at the horizon with the ISCO(Innermost Stable Circular Orbit) and MBCO (Marginally Bound Circular Orbit) for extremal 
Reissner-Nordstr{\o}m(RN) BH and Schwarzschild BH. We find that for extreme RN BH the inequality becomes ${\cal E}_{cm}\mid_{r_{+}}>{\cal E}_{cm}\mid_{r_{mb}}> {\cal E}_{cm}\mid_{r_{ISCO}}$ i.e. ${\cal E}_{cm}\mid_{r_{+}=M}: {\cal E}_{cm}\mid_{r_{mb}=\left(\frac{3+\sqrt{5}}{2}\right)M} : {\cal E}_{cm}\mid_{r_{ISCO}=4M} =\infty : 3.23 : 2.6 $. While for Schwarzschild BH the ratio of CM energy is  ${\cal E}_{cm}\mid_{r_{+}=2M}:  {\cal E}_{cm}\mid_{r_{mb}=4M} : {\cal E}_{cm}\mid_{r_{ISCO}=6M} = \sqrt{5} : \sqrt{2} : \frac{\sqrt{13}}{3}$. Also for
Gibbons-Maeda-Garfinkle-Horowitz-Strominger (GMGHS) BHs,  the ratio is being
${\cal E}_{cm}\mid_{r_{+}=2M}: {\cal E}_{cm}\mid_{r_{mb}=2M} : {\cal E}_{cm}\mid_{r_{ISCO}=2M}
=\infty : \infty : \infty$.
\end{abstract}

\section{Introduction}

In 2009, Ba\~{n}ados, Silk and West\cite{bsw} (hereafter BSW) proposed an
interesting mechanism  that is when two massive dark matter particles falling from rest into the infinite red-shift surface of  a extreme spinning BH described by the Kerr metric can collides with an infinite high center-of-mass(CM) energy. What we will now call the BSW effect. Hence BHs may act as a natural Planck scale particle accelerators. Whereas in the case of non spinning BH described by the Schwarzschild metric, the
CM energy for colliding particles are  finite\cite{baushev}. In a subsequent paper
\cite{silk}, the authors attempted to compute the flux reaching a distant observer
from the annihilation of weakly interacting massive particles  in the
vicinity of an intermediate mass BH.

Soon after the appearance of this mechanism in the literature, several authors were
criticized this effect from the different perspectives.  Particularly in\cite{berti} and  \cite{jacob}, the authors
showed that to operate the BSW process it is indeed require precisely a extremal BH and
extremal horizon respectively. Any deviation  from extremality, the CM energy is also reduced to the order of $10 m_{0}$ or something less, where $m_{0}$ is the mass of the colliding particles.
Also in \cite{berti} the same authors showed that  there is an astrophysical
limitations i.e. maximal spin, back reaction effect and gravitational
radiation etc. on that CM energy due to the Thorn's limit \cite{thorn}
i.e. $\frac{a}{M}=0.998$ ($M$ is the mass and $a$ is the spin
of the BH).  In \cite{piran}, the authors suggested that using
collisional Penrose process the emitted massive  particles can only be
gain $\sim 30$ percentage of the initial rest energy of the in-falling particles.

Also, Lake \cite{lake} calculated the CM energy at the Cauchy horizon of a static Reissner-Nordstr{\o}m (RN) BH and Kerr BH which is limited. Grib and Pablov \cite{grib} investigated the CM energy using the multiple scattering mechanism.
Also in \cite{grib1}, the same authors computed the CM energy of particle collisions
in the ergo-sphere of the Kerr BH. The collision in the ISCO particles was investigated by the \cite{harada} for Kerr BH. Liu et al. \cite{liu} studied the BSW effect for
Kerr-Taub-NUT(Newman, Unti, Tamburino) space-time and proved that the CM energy depends upon  both the Kerr parameter ($a$) and the NUT  parameter ($n$)  of the space-time. In \cite{li}, the authors proved that the Kerr-AdS  BH space-time could act as a particle accelerator. Studies were done by \cite{zong} for RN-de-Sitter  BH and found that infinite energy in the CM frame near the cosmological horizon. The authors in \cite{said} studied the particle accelerations and collisions in the back ground of a cylindrical BHs. Studies of BSW effect were performed in \cite{patil} for the naked singularity case of different space-times.

In \cite{wei1}, the authors discussed the CM energy for the Kerr-Newman BH.
For Kerr-Sen BH, the CM energy is diverging  also discussed in \cite{wei}.
It was discussed in \cite{zaslav} regarding the BSW process for spherically symmetric
RN BH. The authors in \cite{zhu} showed that general stationary
charged BHs as  particle accelerators.  In \cite{frolov}, the author
demonstrated that a weakly magnetized BH may behave as a particle accelerators.
McWilliams \cite{mc} showed that the BHs are neither particle accelerators nor
dark matter probes. The author in \cite{gala} also showed that the CM energy
in the context of the near horizon geometry of the extremal Kerr BHs and proved
that the CM energy is finite for any value of the particle parameters.
Tursunov et al. \cite{tur} have studied the particle accelerations and collisions in case of black string. Fernando \cite{fernando} has studied the possibility of high CM energy
of two particles colliding near the infinite red-shift surface of a charged BH in
string theory.

The aim of this present work is to show that an analogous effect of particle
collision with a high CM energy is also possible when a BH is
described by the \emph{extremal charged dilation  space-time}. We  choose
various collision points say ISCO, MBCO etc. and  at that point we compute
the CM energy and prove that at the horizon the ${\cal E}_{cm}$ is maximum than the
MBCO and ISCO for extremal RN BH. We find that for extreme RN BH the ratio becomes
${\cal E}_{cm}\mid_{r_{+}=M}: {\cal E}_{cm}\mid_{r_{mb}
=\left(\frac{3+\sqrt{5}}{2}\right)M} : {\cal E}_{cm}\mid_{r_{ISCO}=4M}
=\infty : 3.23 : 2.6 $. Which implies that ${\cal E}_{cm}\mid_{r_{+}}>{\cal E}_{cm}\mid_{r_{mb}}> {\cal E}_{cm}\mid_{r_{ISCO}}$. For our completeness
we also compute the ${\cal E}_{cm}$ for Schwarzschild BH and we find the ratio
as ${\cal E}_{cm}\mid_{r_{+}=2M}:  {\cal E}_{cm}\mid_{r_{mb}=4M} : {\cal E}_{cm}\mid_{r_{ISCO}=6M} = \sqrt{5} : \sqrt{2} : \frac{\sqrt{13}}{3}$.
Finally we show the ratio for GMGHS BH and the ratio becomes
${\cal E}_{cm}\mid_{r_{+}=2M}: {\cal E}_{cm}\mid_{r_{mb}=2M}
:{\cal E}_{cm}\mid_{r_{ISCO}=2M} = \infty : \infty : \infty$.  Which is
quite different from extreme RN BH and Schwarzschild BH.

The paper is organized as follows. In section 2, we shall study the basic
properties of charged dilation BH and in the subsection we shall describe
the complete geodesic structure of it. Section 3 will devoted
to study the CM energy of the dilation BH and we shall prove that the
diverging energy can be obtained from the extreme dilation BH. In section
4, we shall discuss the CM energy of particle collision near the
ISCO of an extremal RN BH.  We also discuss the CM energy of particle
collision near the ISCO of a Schwarzschild BH in section 5. In section 6,
we shall describe the CM energy of particle collision near the horizon of
the  Gibbons-Maeda-Garfinkle-Horowitz-Strominger (GMGHS) BH \cite{gm,ghs}.
Finally in section 7, we have given the conclusions and outlook.

\section{Charged dilation metric and its properties:}

The metric of a static, spherically symmetric charged dilation BH \cite{ghs}
can be written in Schwarzschild like coordinates:
\begin{eqnarray}
ds^2=-{\cal F}(r)dt^{2}+\frac{dr^{2}}{{\cal F}(r)}+ {\cal R}^{2}\left(d\theta^{2}+\sin^{2}\theta d\phi^{2}\right) ~.\label{sph}
\end{eqnarray}
where the function ${\cal F}(r)$ is defined by
\begin{eqnarray}
{\cal F}(r) &=& \left(1-\frac{r_{+}}{r}\right)\left(1-\frac{r_{-}}{r}\right)^{\frac{1-a^{2}}{1+a^{2}}}.
\end{eqnarray}
and
\begin{eqnarray}
{\cal R}^{2}(r) &=& r^{2}\left(1-\frac{r_{-}}{r}\right)^{\frac{2a^{2}}{1+a^{2}}}.
\end{eqnarray}
The associated classical dilation field and gauge potential are
\begin{eqnarray}
e^{2a\phi} &=& {\cal R}^{2}(r)= r^{2}\left(1-\frac{r_{-}}{r}\right)^{\frac{2a^{2}}{1+a^{2}}}, \\
{\cal A}&=& \frac{Q}{r} dt,
\\
F &=& d{\cal A} =-\frac{Q}{r^{2}} dt \wedge dr .
\end{eqnarray}
In these equations, $r_{+}$ and  $r_{-}$  are constants, which are related to the mass
and charge of the BH:
\begin{eqnarray}
M &=& \frac{r_{+}}{2} + \left(\frac{1-a^{2}}{1+a^{2}}\right) \frac{r_{-}}{2} \,\, \mbox{and}
 \,\, Q = \sqrt{\frac{r_{+}r_{-}}{1+a^{2}}}
\end{eqnarray}
where $M$ is defined as the BH mass and $Q$ is the electric charge of the BH. It may
be noted that $Q$ and $a$ are positive.
The horizons of the BH are determined by the function ${\cal F}(r)=0$ which yields
\begin{eqnarray}
r_{+} &=& M+ \sqrt{M^{2}-\left(\frac{2n}{1+n}\right)Q^{2}} ~.\label{hocde} \\
r_{-} &=& \frac{1}{n}\left[M+ \sqrt{M^{2}-\left(\frac{2n}{1+n}\right)Q^{2}}\right]
~.\label{hocdc}
\end{eqnarray}
where $n$ is defined by
\begin{eqnarray}
n &=& \frac{1-a^{2}}{1+a^{2}}.
\end{eqnarray}
Here $r_{+}$ and  $r_{-}$ are called event horizon (${\cal H}^+$)  or outer horizon and Cauchy horizon (${\cal H}^-$)  or
inner horizon respectively.  $r_{+}=r_{-}$ or $M^{2}=\left(\frac{1+n}{2}\right)Q^{2}$
corresponds to the extreme charged dilation BH.

{\bf Case I:}  When $a=0$ or $n=1$, the metric corresponds to RN BH.

{\bf Case II:} When $a=1$ or $n=0$, the metric corresponds to GMGHS BH.

The surface gravity of the dilation BH for both the horizons (${\cal H}^\pm$) are
\begin{eqnarray}
{\kappa}_{+} &=& \frac{1}{2r_{+}}\left(\frac{r_{+}-r_{-}}{r_{+}}\right)^{n}
 \,\, \mbox{and} \, \, {\kappa}_{-} =  0  ~.\label{sgcd}
\end{eqnarray}

The  BH temperature or Hawking temperature of ${\cal H}^\pm$ are
\begin{eqnarray}
T_{+}&=& \frac{\kappa_{+}}{2\pi} \\
 &=& \frac{1}{4\pi r_{+}}\left(\frac{r_{+}-r_{-}}{r_{+}}\right)^{n} \\
\mbox{and} \nonumber\\
T_{-}&=& \frac{\kappa_{-}}{2\pi}=0
\end{eqnarray}

Then the area of both the horizons (${\cal H}^\pm$) are
\begin{eqnarray}
{\cal A}_{+} &=& 4\pi {\cal R}_{+}^{2} =4\pi r_{+}^2 \left(\frac{r_{+}-r_{-}}{r_{+}}\right)^{1-n} \,\, \mbox{and} \, \, {\cal A}_{-}=4\pi {\cal R}_{-}^{2} = 0\,\, ~.\label{arcd}
\end{eqnarray}
Interestingly, the area of both the horizons go  to zero at the extremal limit ($r_{+}=r_{-}$). This feature is quite different from the well known RN and Schwarzschild BH. The other characteristics of this space-time is that there is a curvature singularity at $r=r_{-}$.

Finally, the entropy of both the horizons (${\cal H}^\pm$) of the BH reads as
\begin{eqnarray}
{\cal S}_{+} &=& \frac{{\cal A}_{+}}{4} = \pi r_{+}^2 \left(\frac{r_{+}-r_{-}}{r_{+}}\right)^{1-n} \,\, \mbox{and}\,\,  {\cal S}_{-}= \frac{{\cal A}_{-}}{4}= 0\,\,~.\label{etpcd}
\end{eqnarray}

\subsection{Equatorial circular orbit in the charged dilation BH:}

This section is devoted to study the properties of the circular geodesics
for charged dilation BH in the $\theta=\frac{\pi}{2}$ plane and also compute
the ISCO equation for this BH.

To compute the geodesic motion of a test particle in the equatorial plane
we set $u^{\theta}=\dot{\theta}=0$ and $\theta=constant=\frac{\pi}{2}$ and
follow the pioneer book of S. Chandrasekhar\cite{sch}.
Thus we may write the Lagrangian density in terms of the metric is given by
\begin{eqnarray}
2{\cal L} &=& -{\cal F}(r)\,(u^{t})^2+({\cal F}(r))^{-1}\,
(u^{r})^2 +{\cal R}^{2}(r)\,(u^{\phi})^2 ~.\label{lags}
\end{eqnarray}
The generalized momenta reads
\begin{eqnarray}
p_{t} & \equiv & \frac{\partial {\cal L}}{\partial \dot{t}}=-{\cal F}(r)\,u^{t} ~.\label{pts}\\
p_{r} & \equiv & \frac{\partial {\cal L}}{\partial \dot{r}}= ({\cal F}(r))^{-1}\, u^{r}  ~.\label{prs}\\
p_{\theta} & \equiv & \frac{\partial {\cal L}}{\partial \dot{\theta}} ={\cal R}^{2}(r) \,u^{\theta}=0  ~.\label{prtheta}\\
p_{\phi} & \equiv & \frac{\partial {\cal L}}{\partial \dot{\phi}} = {\cal R}^{2}(r) \,u^{\phi} ~.\label{pps}
\end{eqnarray}
Here superior dots denote differentiation with respect to affine parameter
which is the proper time ($\tau$) for time-like case and for null it is
affine parameter($\lambda$).

Since the Lagrangian density does not depend explicitly on the variables
`$t$' and `$\phi$', so $p_{t}$ and $p_{\phi}$ are conserved quantities.
Thus one gets,
\begin{eqnarray}
p_{t} &= &-{\cal F}(r)\,u^{t} =- {\cal E}=constant~.\label{pts1}\\
p_{\phi} & =& {\cal R}^{2}(r) \,u^{\phi}= L =constant ~.\label{pps1}
\end{eqnarray}

Solving (\ref{pts1}) and (\ref{pps1}) for $u^{t}$ and $u^{\phi}$ we obtain
\begin{eqnarray}
u^{t}=\frac{{\cal E}}{{\cal F}(r)} ~.\label{tdot} \\
u^{\phi}=\frac{L}{{\cal R}^{2}(r)}~.\label{phid}
\end{eqnarray}
where ${\cal E}$ and $L$ are the energy per unit mass  and angular momentum per unit
mass of the test particle.

Therefore the required Hamiltonian reads as
\begin{eqnarray}
{\cal H} &=& p_{t}\,u^{t}+p_{\phi}\,u^{\phi}+p_{r}\,u^{r}-\cal L ~.\label{hams}
\end{eqnarray}

In terms of the metric the Hamiltonian  may be written as
\begin{eqnarray}
{\cal H} &=& -{\cal F}(r)\,(u^{t})^{2}+({\cal F}(r))^{-1}(u^{r})^2+{\cal R}^{2}(r)(u^\phi)^2-{\cal L} ~.\label{hams1}
\end{eqnarray}
Since the Hamiltonian is independent of $`t'$, therefore we can write it as
\begin{eqnarray}
2{\cal H} &=& -{\cal F}(r)\,(u^{t})^{2}+({\cal F}(r))^{-1}(u^{r})^2+{\cal R}^{2}(r)(u^\phi)^2 ~.\label{ham1}\\
&=&-{\cal E}\,u^{t}+ L \,u^{\phi}+\frac{1}{{\cal F}(r)}\,(u^r)^2=\epsilon=const ~.\label{hs1}
\end{eqnarray}
Here $\epsilon=-1$ for time-like geodesics, $\epsilon=0$ for light-like geodesics and $\epsilon=+1$ for space-like geodesics.

Substituting the equations, (\ref{tdot}) and (\ref{phid}) in (\ref{hs1}), we find the radial equation for charged dilation BH:
\begin{eqnarray}
(u^{r})^{2}=\dot{r}^{2}={\cal E}^{2}-{\cal V}_{eff}={\cal E}^{2}-\left(\frac{L^{2}}{{\cal R}^{2}(r)}-\epsilon \right){\cal F}(r) ~.\label{radial}
\end{eqnarray}
where, the standard effective potential for charged dilation  space-time becomes
\begin{eqnarray}
{\cal V}_{eff}=\left(\frac{L^{2}}{{\cal R}^{2}(r)}-\epsilon \right){\cal F}(r) ~.\label{vrn}
\end{eqnarray}

\subsection{Time-like Case:}

For time-like case the effective potential is found to be
\begin{eqnarray}
{\cal V}_{eff}= \left(1-\frac{r_{+}}{r}\right)\left(1-\frac{r_{-}}{r}\right)^{n}
\left[1+\frac{L^{2}}{r^{2}}\left(1-\frac{r_{-}}{r}\right)^{n-1}\right] ~.\label{vrt}
\end{eqnarray}
In the extremal case, the effective potential for charged dilation BH reduced to
the following form:
\begin{eqnarray}
{\cal V}_{eff}= \left(1-\frac{r_{+}}{r}\right)^{n+1}+
\frac{L^{2}}{r^{2}}\left(1-\frac{r_{+}}{r}\right)^{2n} ~.\label{vrtx}
\end{eqnarray}

For circular geodesic motion of the test particle of constant $r=r_{0}$, we must have
\begin{eqnarray}
\dot{r} &=& 0 \,\, \mbox{or} \,\, {\cal V}_{eff} = {\cal E}^{2} ~.\label{v}
\end{eqnarray}
and
\begin{eqnarray}
\frac{d{\cal V}_{eff}}{dr} &=& 0 ~.\label{dvdr}
\end{eqnarray}

Thus one can obtain the conserved energy and angular momentum per unit mass of the
test particle along the circular orbits are:
\begin{eqnarray}
{\cal E}^{2}_{0} &=& \frac{r_{0}\left(1-\frac{r_{+}}{r_{0}}\right)^{2}\left(1-\frac{r_{-}}{r_{0}}\right)^{n}
\left[2r_{0}\left(1-\frac{r_{-}}{r_{0}}\right)+r_{-}(1-n)\right]}
{\left[2r_{0}^{2}-(3r_{+}+(2n+1)r_{-})r_{0}+2(1+n)r_{+}r_{-}\right]} ~.\label{engcd}
\end{eqnarray}
and
\begin{eqnarray}
L_{0}^{2} &=& \frac{r_{0}^{3}\left(1-\frac{r_{-}}{r_{0}}\right)^{1-n}
\left[r_{+}\left(1-\frac{r_{-}}{r_{0}}\right)+nr_{-}\left(1-\frac{r_{+}}{r_{0}}\right)\right]}
{\left[2r_{0}^{2}-(3r_{+}+(2n+1)r_{-})r_{0}+2(1+n)r_{+}r_{-}\right]}~ .\label{angcd}
\end{eqnarray}

{\bf Case I:}  When  $n=1$, we retrieve the value of energy and angular momentum
 of RN BH. \\
{\bf Case II:} When $n=0$, we retrieve the value of energy and angular momentum
 of  GMGHS BH.

Also in the extremal limit, the energy and angular momentum corresponds to
\begin{eqnarray}
{\cal E}^{2}_{0} &=& \frac{r_{0}\left(1-\frac{r_{+}}{r_{0}}\right)^{n+2}
\left[2r_{0}-(1+n)r_{+}\right]}
{2\left[r_{0}^{2}-(n+2)r_{+}r_{0}+(1+n)r_{+}^{2}\right]} ~.\label{engcdx}
\end{eqnarray}
and
\begin{eqnarray}
L_{0}^{2} &=& \frac{(n+1)r_{+}r_{0}^{3}\left(1-\frac{r_{+}}{r_{0}}\right)^{2-n}}
{2\left[r_{0}^{2}-(n+2)r_{+}r_{0}+(1+n)r_{+}^{2}\right]} ~.\label{angcdx}
\end{eqnarray}

{\bf Case I:}  When  $n=1$, we retrieve the value of energy and angular momentum
 of extreme RN BH. \\
{\bf Case II:} When $n=0$, we retrieve the value of energy and angular momentum
 of extreme  GMGHS BH.

Circular motion of the test particle to be exists when both the energy and angular
momentum are real and finite. Therefore  we must have
\begin{eqnarray}
2r_{0}^{2}-(3r_{+}+(2n+1)r_{-})r_{0}+2(1+n)r_{+}r_{-} > 0 \,\, \mbox{and} \,\, r_{0}>r_{-}
~. \label{crcd}
\end{eqnarray}
General relativity does not permit arbitrary circular radii, so the denominator of equations
(\ref{engcd},\ref{angcd}) real only if $2r_{0}^{2}-(3r_{+}+(2n+1)r_{-})r_{0}+2(1+n)r_{+}r_{-}\geq 0$. The
limiting case of equality gives an circular orbit with indefinite energy
per unit mass, i.e. a  circular photon orbit (CPO). This photon orbit is
 the innermost boundary of the circular orbit for massive particles.

Comparing the above equation of particle orbits with (\ref{ph1cd}) when
$r_{0}=r_{c}$, we can observe that photon orbits are the limiting case
of time-like circular orbit.  It occurs at the radius
\begin{eqnarray}
r_{c} &=&  r_{ph}= \frac{1}{4}\left[\left(3r_{+}+(2n+1)r_{-}\right)\pm \sqrt{\left(3r_{+}+(2n+1)r_{-}\right)^{2}
 -16(1+n)r_{+}r_{-}} \right]
\end{eqnarray}
In the extremal limit,  this equation gives
\begin{eqnarray}
 r_{ph} &=& (n+1)r_{+}
\end{eqnarray}
When $n=0$, we recover the value of CPO of GMGHS BH which is
$r_{ph}=r_{+}=2M$ and when $n=1$, we recover the value of CPO of RN BH which is $r_{ph}=2r_{+}=2M$.

\subsection{Circular Photon Orbits:}

For null circular geodesics, the effective potential becomes
\begin{eqnarray}
{\cal U}_{eff} &=& \frac{L^2}{{\cal R}^{2}(r)}{\cal F}(r)
=\frac{L^2}{r^{2}}\left(1-\frac{r_{+}}{r}\right)\left(1-\frac{r_{-}}{r}\right)^{2n-1}
\end{eqnarray}
In the extremal limit, the effective potential goes to the following form:
\begin{eqnarray}
{\cal U}_{eff} &=& \frac{L^2}{{\cal R}^{2}(r)}{\cal F}(r)
=\frac{L^2}{r^{2}}\left(1-\frac{r_{+}}{r}\right)^{2n}
\end{eqnarray}
For circular null geodesics at $r=r_{c}$, we find
\begin{eqnarray}
{\cal U}_{eff} &=& {\cal E}^2
\end{eqnarray}
and
\begin{eqnarray}
 \frac{d{\cal U}_{eff}}{dr} &=& 0
\end{eqnarray}

Thus one may  obtain the ratio of energy and angular momentum of the test particle
evaluated at $r=r_{c}$ for circular photon orbits are:

\begin{eqnarray}
\frac{{\cal E}_{c}}{L_{c}} &=& \sqrt{\frac{\left(1-\frac{r_{+}}{r_{c}}\right)\left(1-\frac{r_{-}}{r_{c}}\right)^{2n-1}} {r_{c}^{2}}}
\end{eqnarray}
and
\begin{eqnarray}
{\cal R}(r_{c}){\cal F}'(r_{c})
   -2{\cal F}(r_{c}){\cal R}'(r_{c}) &=& 0 ~.\label{nulcd}
\end{eqnarray}
or
\begin{eqnarray}
2r_{c}^{2}-(3r_{+}+(2n+1)r_{-})r_{c}+2(1+n)r_{+}r_{-}  &=& 0 .~\label{ph1cd}
\end{eqnarray}

Let $r_{c}=r_{ph}$ be the  solution of the equation (\ref{ph1cd}) which gives
the radius of the circular photon orbit (CPO) of the  charged dilation space-time.

{\bf Case I:}  When  $n=1$, we recover the CPO of RN BH. \\
{\bf Case II:} When $n=0$, we retrieve the CPO of  GMGHS BH.
After incorporating the impact parameter $D_{c}=\frac{L_{c}}{E_{c}}$, the above
equation can be written as
\begin{eqnarray}
 \frac{1}{D_{c}} &=& \frac{{\cal E}_{c}}{L_{c}}
=\sqrt{\frac{\left(1-\frac{r_{+}}{r_{c}}\right)\left(1-\frac{r_{-}}
 {r_{c}}\right)^{2n-1}}{r_{c}^{2}}}
\end{eqnarray}
At the extremal limit $r_{+}=r_{-}$, this would be
\begin{eqnarray}
 \frac{1}{D_{c}} &=& \frac{{\cal E}_{c}}{L_{c}}
=\frac{\left(1-\frac{r_{+}}{r_{c}}\right)^{n}}{r_{c}}
\end{eqnarray}

The classical capture cross-section is given by
\begin{eqnarray}
\sigma &=& \pi D_{c}^{2}=\frac{r_{c}^{2}\left(1-\frac{r_{-}}{r_{c}}\right)^{1-2n}}
{\left(1-\frac{r_{+}}{r_{c}}\right)}
\end{eqnarray}

{\bf Case I:}  When  $n=1$, we recover the classical capture cross-section
 of RN BH. \\
{\bf Case II:} When $n=0$, we retrieve the classical capture cross-section
 of  GMGHS BH.

The another important class orbit is the marginally bound circular orbit (MBCO)
can be found by setting ${\cal E}_{0}^{2}=1$  in Eq. (\ref{engcd}), then the MBCO
equation reads as
$$
2\left[1-\left(1-\frac{r_{+}}{r_{0}}\right)^{2}
\left(1-\frac{r_{-}}{r_{0}}\right)^{n+1}\right]r_{0}^{2} -
$$
\begin{eqnarray}
\left[3r_{+}+\{(2n+1)-(1-n)\left(1-\frac{r_{+}}{r_{0}}\right)^{2}
\left(1-\frac{r_{-}}{r_{0}}\right)^{n}\}r_{-}\right]r_{0}+2(1+n)r_{+}r_{-} &=& 0
~. \label{mbcocd}
\end{eqnarray}

{\bf Case I:}  When  $n=1$, we retrieve  the MBCO of RN BH,  which may be determined
from the following equation:
\begin{eqnarray}
(r_{+}+r_{-})r_{0}^{3}-2(r_{+}+r_{-})^{2}r_{0}^{2}+4r_{+}r_{-}(r_{+}+r_{-})r_{0}
-2(r_{+}r_{-})^{2} &=& 0
\end{eqnarray}
Let $r_{0}=r_{mb}$ be the solution of the equation which gives the radius of
MBCO of RN BH.

{\bf Case II:}
When $n=0$, we retrieve the MBCO of  GMGHS BH,  which can be
determined from the following equation:
\begin{eqnarray}
r_{0}^{2}-2r_{+}r_{0}+r_{+}r_{-} &=& 0
\end{eqnarray}
or
\begin{eqnarray}
r_{0} &=& r_{mb}= r_{+}\pm \sqrt{r_{+}(r_{+}-r_{-})}
\end{eqnarray}

From the astrophysical point of view, the most important class of orbit
is the ISCO which may be derived from the second derivative of the
effective potential of time-like case.
i.e.
\begin{eqnarray}
\frac{d^2{\cal V}_{eff}}{dr^2} &=& 0 \label{pincd}
\end{eqnarray}
Thus one may obtain the ISCO equation for the charge dilation space-time
from the following functional equation:
$$
{\cal F}(r_{0}) {\cal F}''(r_{0}){\cal R}(r_{0}){\cal R}'(r_{0})-{\cal F}(r_{0})
{\cal F}'(r_{0}){\cal R}(r_{0}){\cal R}''(r_{0})-
$$
\begin{eqnarray}
2{\cal R}(r_{0}){\cal R}'(r_{0})({\cal F}'(r_{0}))^{2}
   -3{\cal F}(r_{0}) {\cal F}'(r_{0})({\cal R}'(r_{0}))^{2} &=& 0 ~.\label{iscocd}
\end{eqnarray}
For extremal case $(r_{+}=r_{-})$, it can be easily obtained the ISCO equation for
charged dilation BH which is given by
\begin{eqnarray}
r_{0}^{2}-(3n+2)r_{+}r_{0}+(n+1)^{2}r_{+}^{2} &=& 0
\end{eqnarray}
or
\begin{eqnarray}
r_{0} &=& \frac{r_{+}}{2}\left[(3n+2)\pm \sqrt{n(5n+4)}\right]
\end{eqnarray}

In the limit $ n \rightarrow 1$, we obtain the radius of ISCO for RN BH,  which
is $r_{0}=r_{ISCO}=4M$ and in the limit $ n \rightarrow 0$, we recover the radius
of ISCO for GMGHS BH which is $r_{0}=r_{ISCO}=2M$.

\section{CM Energy of Particle Collision near the horizon of the Dilation BH:}

This section is devoted to study the particle acceleration and collision in the
CM  frame. To compute the CM energy, we consider two particles coming from
infinity with $\frac{{\cal E}_{1}}{m_{0}}=\frac{{\cal E}_{2}}{m_{0}}=1$
approaching the charged dilation BH with different angular momenta $L_{1}$ and $L_{2}$
and colliding at some radius $r$. Later, we consider the collision point $r$
to approach the horizon $r=r_{+}$. Also we have assumed the the particles to
be at rest at infinity.

The CM energy is evaluated by using the following formula which
was first given  by BSW \cite{bsw} reads
\begin{eqnarray}
\left(\frac{{\cal E}_{cm}}{\sqrt{2}m_{0}}\right)^{2} &=&
1-g_{\mu\nu}u^{\mu}_{1}u^{\nu}_{2}~.\label{cm}
\end{eqnarray}

We shall also assume throughout this work the geodesic motion of the colliding particles
confined in the $\theta=\frac{\pi}{2}$ plane. Since  the charged dilation
 space-time has a time-like isometry followed by the
time-like Killing vector field $\chi$ whose projection along the four velocity
${\bf u}$ of geodesics $\chi.{\bf u}=-{\cal E}$, is conserved along such geodesics
(where $\chi\equiv \partial_{t}$). Similarly  there is also the `angular momentum'
$L=\zeta.{\bf u}$ is conserved due to the rotational symmetry
(where $\zeta\equiv \partial_{\phi})$.

For massive particles, the components of the four velocity are
\begin{eqnarray}
  u^{t} &=& \dot{t} =\frac{{{\cal E}}}{{\cal F}(r)}  \\
  u^{r} &=& \dot{r}=\pm \sqrt{{\cal E}^{2}-{\cal F}(r)
          \left(1+\frac{L^{2}}{{\cal R}^{2}(r)}\right)}\label{eff}\\
  u^{\theta} &=& \dot{\theta} = 0 \\
  u^{\phi} &=& \dot{\phi} = \frac{L}{{\cal R}^{2}(r)} ~.\label{utur}
\end{eqnarray}
and

\begin{eqnarray}
u^{\mu}_{1} &=& \left( \frac{{\cal E}_{1}}{{\cal F}(r)},~ -X_{1},~ 0,~\frac{L_{1}}{{\cal R}^{2}(r)}\right) ~.\label{u1}\\
u^{\mu}_{2} &=& \left( \frac{{\cal E}_{2}}{{\cal F}(r)},~ -X_{2},~ 0,~\frac{L_{2}}{{\cal R}^{2}(r)}\right) ~.\label{u2}
\end{eqnarray}

Substituting this in (\ref{cm}), we get the center of mass energy:
\begin{eqnarray}
\left(\frac{{\cal E}_{cm}}{\sqrt{2}m_{0}}\right)^{2} &=&  1 +\frac{{\cal E}_{1}
{\cal E}_{2}}{{\cal F}(r)}-
\frac{X_{1}X_{2}}{{\cal F}(r)}-\frac{L_{1}L_{2}}{{\cal R}^{2}(r)} ~.\label{cm1}
\end{eqnarray}
where,
$$
X_{1} = \sqrt{{\cal E}_{1}^{2}-{\cal F}(r)\left(1+\frac{L_{1}^{2}}{{\cal R}^{2}(r)}\right)}, \,\,
\\
X_{2} = \sqrt{{\cal E}_{2}^{2}-{\cal F}(r)\left(1+\frac{L_{2}^{2}}{{\cal R}^{2}(r)}\right)}
$$

For simplicity, ${\cal E}_{1}={\cal E}_{2}=1$ and substituting the value of ${\cal F}(r)$,
we obtain the CM energy near the event horizon ($r_{+}$) of the charged dilation space-time:
\begin{eqnarray}
{\cal E}_{cm}\mid_{r\rightarrow r_{+}} &=& \sqrt{2}m_{0}
\sqrt{\frac{4{\cal R}_{+}^2+(L_{1}-L_{2})^{2}}{2{\cal R}_{+}^2}} ~.\label{cmcd}
\end{eqnarray}
where $r_{+}$ is described in equation (\ref{hocde}). Putting the value of
${\cal R}_{+}^2=r_{+}^{1+n}(r_{+}-r_{-})^{1-n}$ in the above equation  we
get
\begin{eqnarray}
{\cal E}_{cm}\mid_{r\rightarrow r_{+}} &=& \sqrt{2}m_{0}
\sqrt{\frac{4{r}_{+}^{n+1}(r_{+}-r_{-})^{1-n}+(L_{1}-L_{2})^{2}}
{2{r}_{+}^{n+1}(r_{+}-r_{-})^{1-n}}} ~.\label{cmcd1}
\end{eqnarray}
Which shows that the CM energy is finite and depends upon the values of the
angular momentum parameter.

Whenever we taking the extremal limit $r_{+}=r_{-}$, we get the CM energy near the
event horizon $r=r_{+}$:
\begin{eqnarray}
{\cal E}_{cm} \mid_{r\rightarrow r_{+}} &\longmapsto &  \infty
\end{eqnarray}
which suggests that the CM energy of collision for extremal charged
dilation BH is diverging  as we approaches the extremal limit. Thus we get the
unlimited CM energies. Interestingly, it is independent of the fine tuning
of the angular momentum parameter. This is one of the key point of the paper.

{\bf Case I:}  When  $n=1$, we recover the value of CM energy of RN BH, which
is given by
\begin{eqnarray}
{\cal E}_{cm}\mid_{r\rightarrow r_{+}} &=&
\sqrt{2}m_{0}\sqrt{\frac{4r_{+}^2+(L_{1}-L_{2})^{2}}{2r_{+}^2}} ~.\label{cm5}
\end{eqnarray}

{\bf Case II:} When $n=0$, we retrieve the  value of CM energy of  GMGHS BH
which is given by
\begin{eqnarray}
{\cal E}_{cm} &=& \sqrt{2}m_{0}\sqrt{\frac{8M(2M-b)+(L_{1}-L_{2})^{2}}{4M(2M-b)}}
\end{eqnarray}

In the limit $b\rightarrow 0$, the above expression reduces to
\begin{eqnarray}
{\cal E}_{cm} &=& \sqrt{2}m_{0}\sqrt{\frac{16M^{2}+(L_{1}-L_{2})^{2}}{8M^{2}}}~.\label{cmsch}
\end{eqnarray}
which is the CM energy of the Schwarzschild BH. In fact this
is indeed a finite quantity first observed in \cite{bsw}.

Interestingly, near the Cauchy horizon $(r_{-})$ the CM energy for
charged dilation space-time is found to  be
\begin{eqnarray}
{\cal E}_{cm}\mid_{r\rightarrow r_{-}} &=& \sqrt{2}m_{0}
\sqrt{\frac{4{\cal R}_{-}^2+(L_{1}-L_{2})^{2}}{2{\cal R}_{-}^2}} ~.\label{cmcd2}
\end{eqnarray}
where $r_{-}$ is described in equation (\ref{hocdc}).

Since the values of ${\cal R}_{-}^2$ is zero, i.e. ${\cal R}_{-}^2=0$, thus the value of CM
energy near the inner horizon leads to diverging value.
Hence,
\begin{eqnarray}
{\cal E}_{cm}\mid_{r\rightarrow r_{-}}  &\longmapsto &  \infty
\end{eqnarray}
This is another interesting feature of the charged dilation BH.

\section{CM Energy of Particle Collision near the ISCO of an Extremal RN BH:}

The aim of this section is to compute the CM energy for RN BH at $r=r_{ISCO}$,
$r=r_{mb}$ and $r=r_{+}$. Then we compare the results obtained for the different
collision point.
The metric for RN space-time can be written as

\begin{eqnarray}
ds^2=-{\cal G}(r)dt^{2}+\frac{dr^{2}}{{\cal G}(r)}+ r^{2}\left(d\theta^{2}+\sin^{2}\theta d\phi^{2}\right) ~.\label{sprn}
\end{eqnarray}
where the function ${\cal G}(r)$ is defined by
\begin{eqnarray}
{\cal G}(r) &=& \left(1-\frac{r_{+}}{r}\right)\left(1-\frac{r_{-}}{r}\right).
\end{eqnarray}
The BH has event horizon which is situated at $r_{+}=M+\sqrt{M^{2}-Q^{2}}$ and
Cauchy horizon which is situated at $r_{-}=M-\sqrt{M^{2}-Q^{2}}$.
The equatorial time-like geodesics of RN BHs are (See Figs. 1-2)
\begin{eqnarray}
  u^{t} &=& \frac{{{\cal E}}}{{\cal G}(r)}  \\
  u^{r} &=& \pm \sqrt{{\cal E}^{2}-{\cal G}(r)\left(1+\frac{L^{2}}{r^2}\right)} \\
  u^{\theta} &=& 0 \\
  u^{\phi} &=& \frac{L}{r^2} ~.\label{utur1}
\end{eqnarray}





\begin{figure}[t]
\begin{center}
\subfigure[][]{\includegraphics[width=0.45\textwidth]{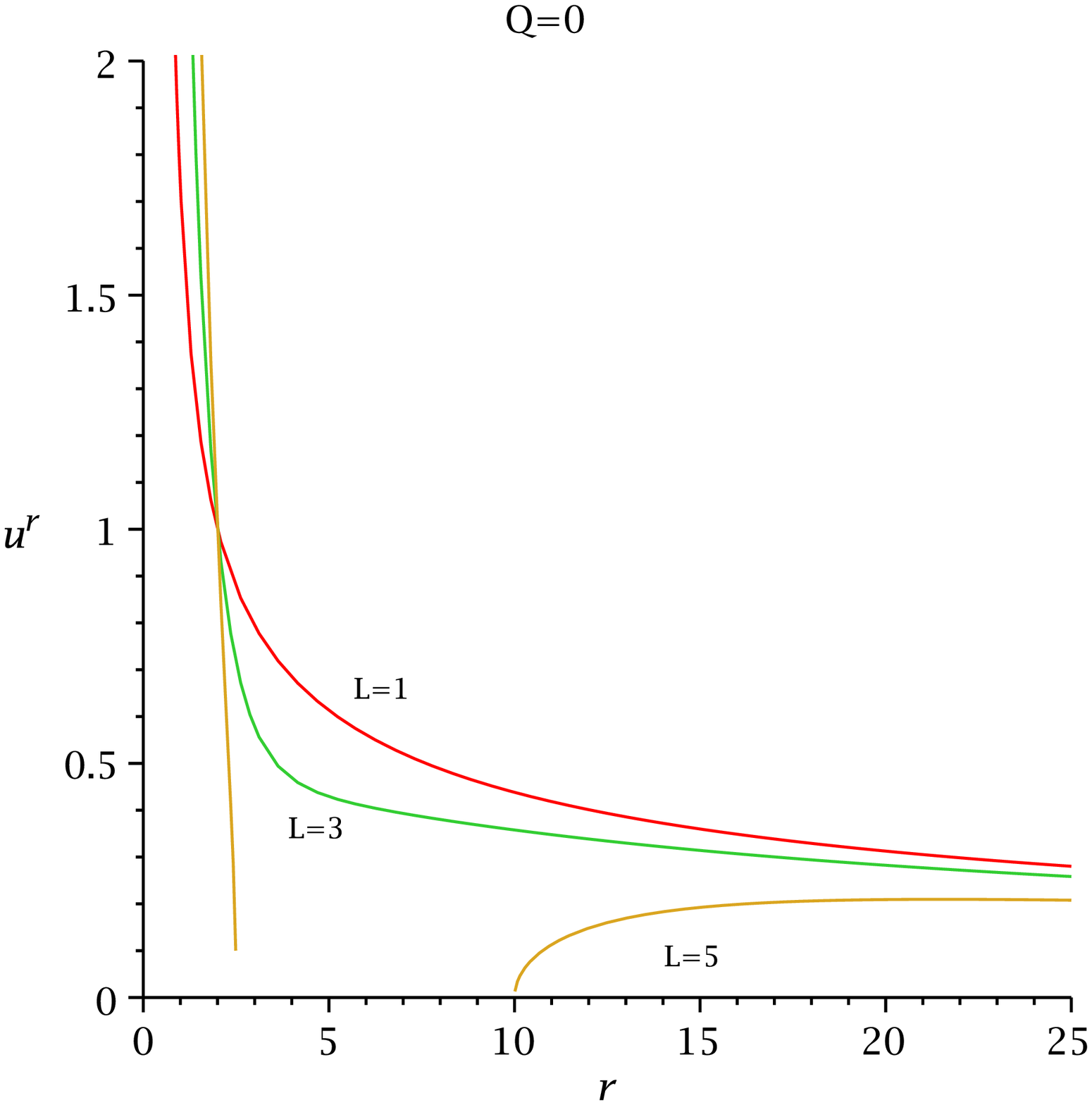}}\qquad
\subfigure[][]{\includegraphics[width=0.45\textwidth]{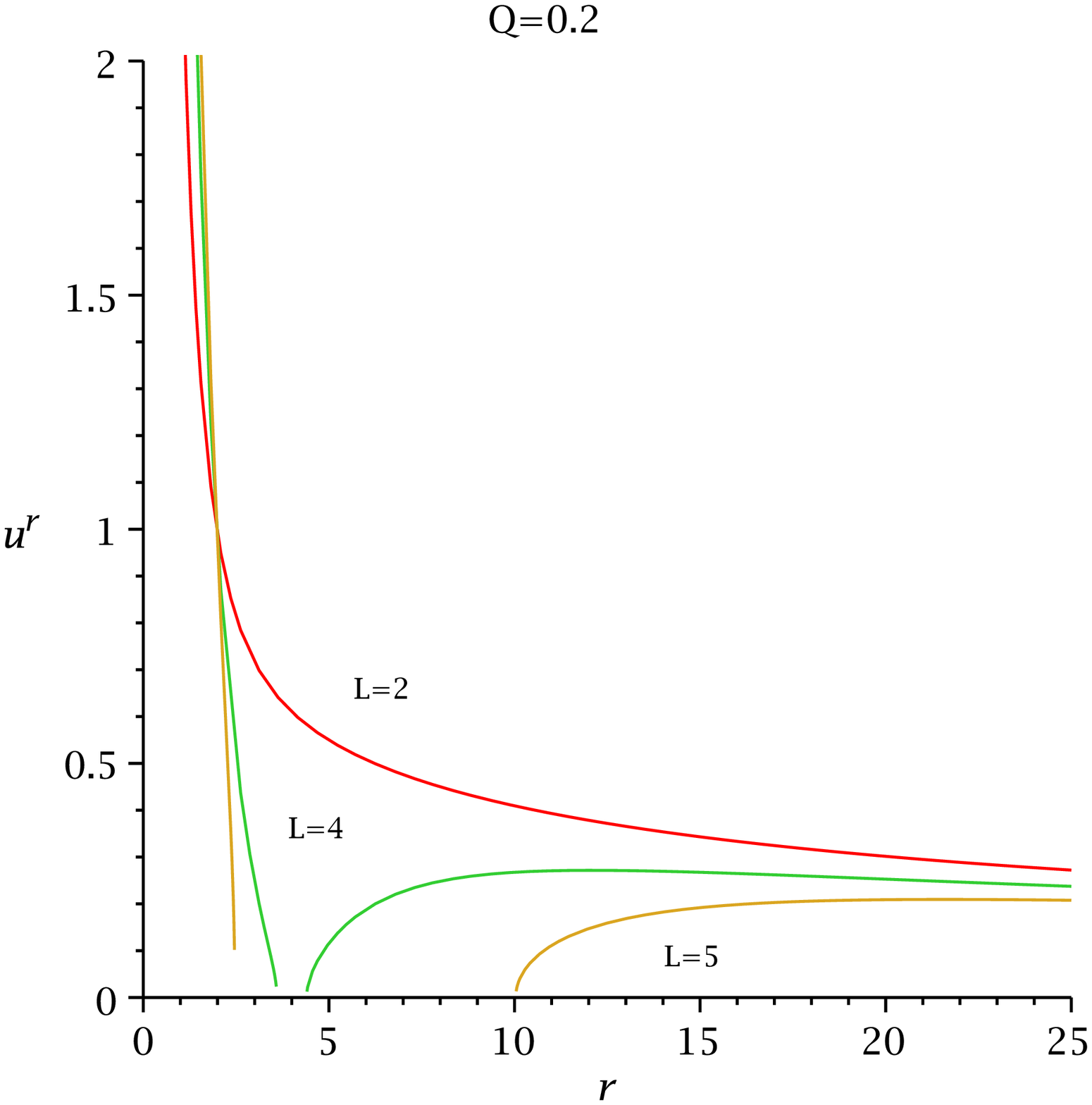}}
\end{center}
\caption{The figure shows the variation  of $\dot{r}$  with $r$ for RN BH with $Q=0$
and $Q=0.2$.
 \label{rncase}}
\end{figure}

\begin{figure}[t]
\begin{center}
\subfigure[][]{\includegraphics[width=0.45\textwidth]{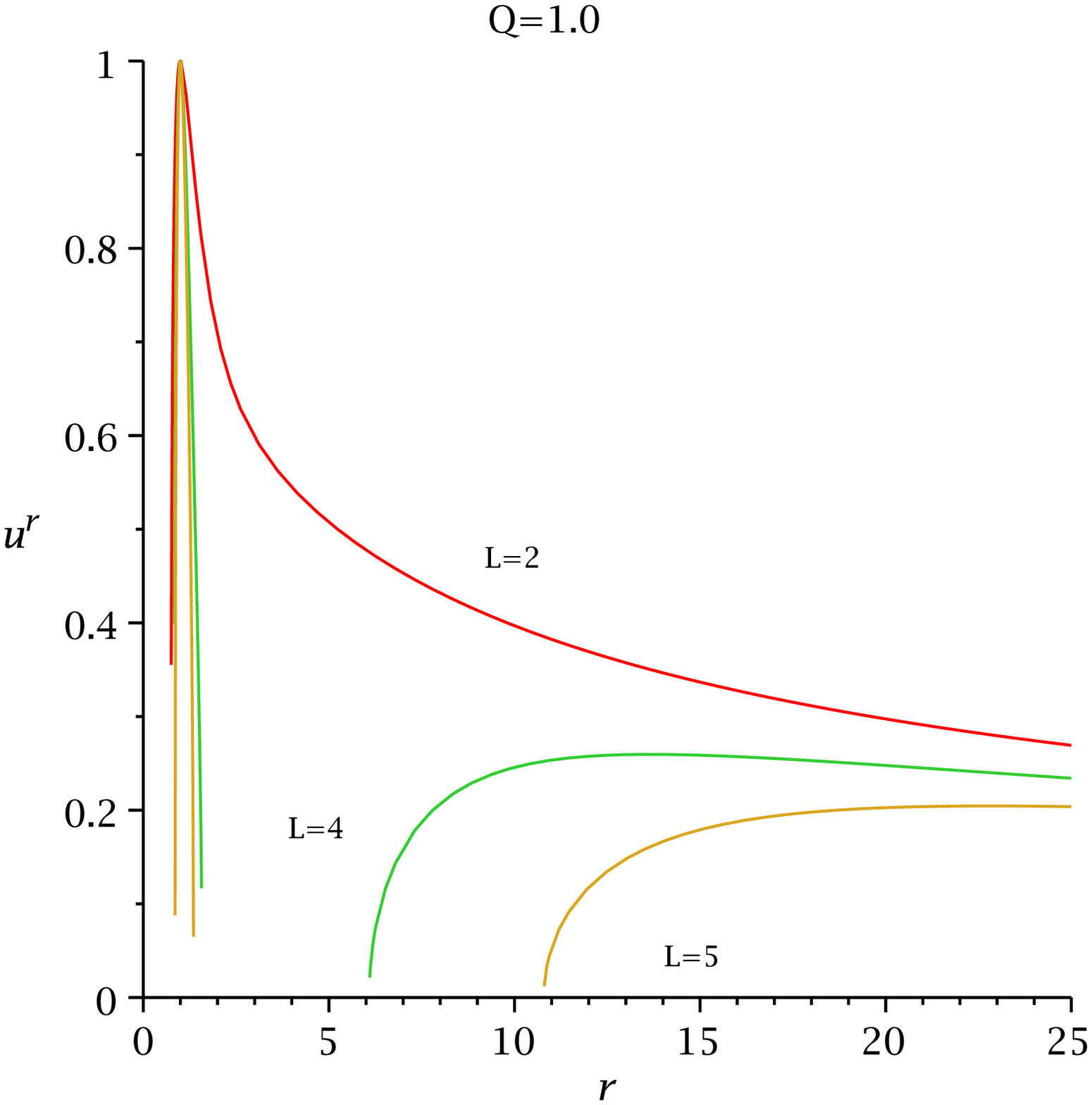}}\qquad
\subfigure[][]{\includegraphics[width=0.45\textwidth]{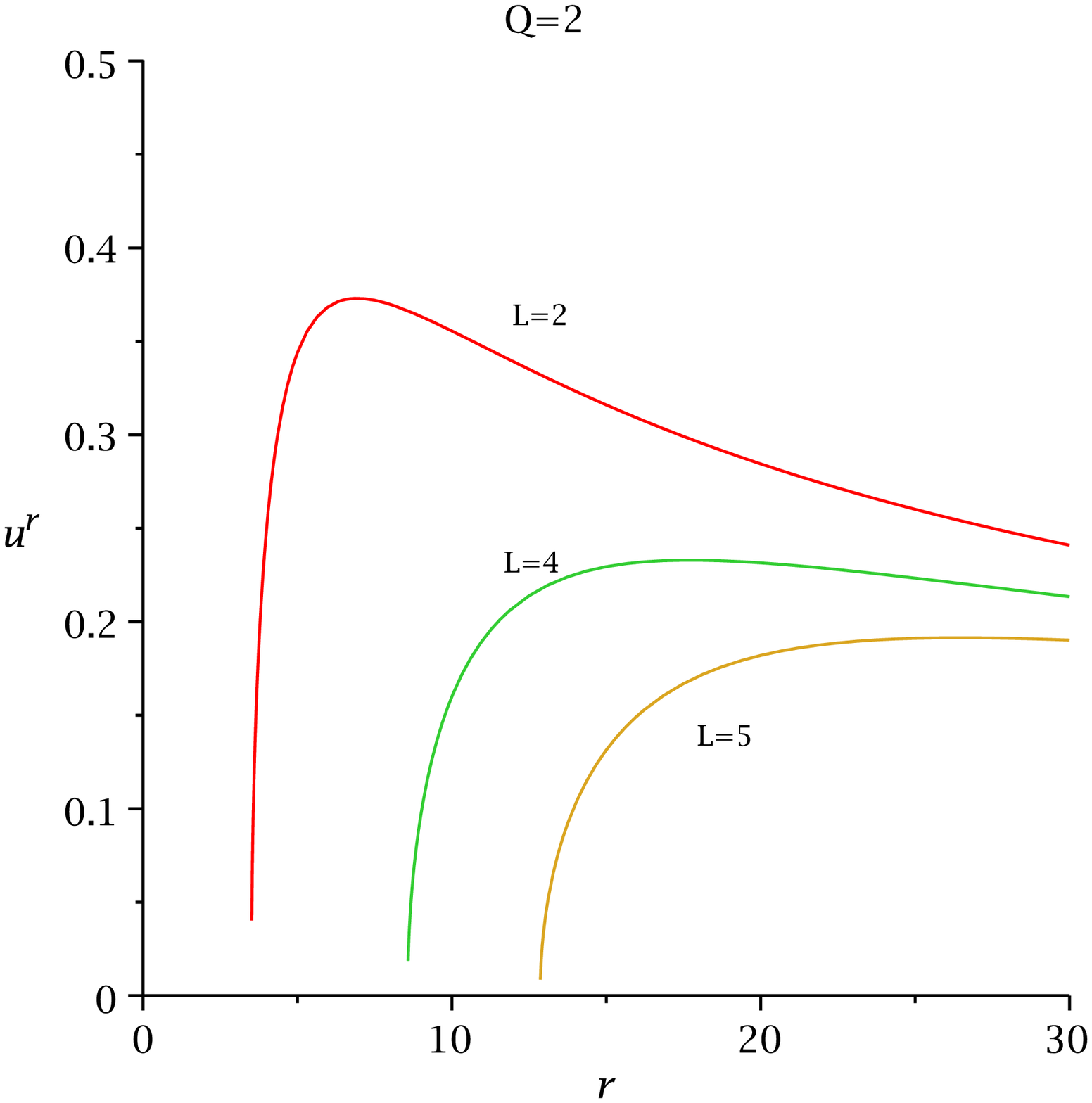}}
\end{center}
\caption{The figure shows the variation  of $\dot{r}$  with $r$ for RN BH with $Q=1.0$
and $Q=2.0$.
 \label{rncase1}}
\end{figure}

Next, we compute the CM energy for the RN space-time by using the well known
formula as given by  Ba\~{n}ados et al. in \cite{bsw}:

\begin{eqnarray}
\left(\frac{{\cal E}_{cm}}{\sqrt{2}m_{0}}\right)^{2} &=&  1-g_{\mu\nu}u^{\mu}_{1}u^{\nu}_{2}~.\label{cm3}
\end{eqnarray}
where $u^{\mu}_{1}$ and $u^{\nu}_{2}$ are the 4-velocity of the two particles, which can be found from the following equation(\ref{utur1}).

\begin{eqnarray}
u^{\mu}_{1} &=&
\left( \frac{{\cal E}_{1}}{{\cal G}(r)},~ -Y_{1},~ 0,~\frac{L_{1}}{r^{2}}\right)
~.\label{u3}\\
u^{\mu}_{2} &=&
\left( \frac{{\cal E}_{2}}{{\cal G}(r)},~ -Y_{2},~ 0,~\frac{L_{2}}{r^{2}}\right)
~.\label{u4}
\end{eqnarray}

Therefore using(\ref{cm3}) one can obtain the center-of-mass energy for this collision:
\begin{eqnarray}
\left(\frac{{\cal E}_{cm}}{\sqrt{2}m_{0}}\right)^{2} &=&
 1 +\frac{{\cal E}_{1}{\cal E}_{2}}{{\cal G}(r)}
-\frac{Y_{1}Y_{2}}{{\cal G}(r)}-\frac{L_{1}L_{2}}{r^2}~.\label{cm4}\\
\mbox{where}\\ \nonumber
Y_{1} &=& \sqrt{{\cal E}_{1}^{2}-{\cal G}(r)\left(1+\frac{L_{1}^{2}}{r^2}\right)}\\
Y_{2} &=& \sqrt{{\cal E}_{2}^{2}-{\cal G}(r)\left(1+\frac{L_{2}^{2}}{r^2}\right)}
\end{eqnarray}




\begin{figure}[t]
\begin{center}
\subfigure[][]{\includegraphics[width=0.45\textwidth]{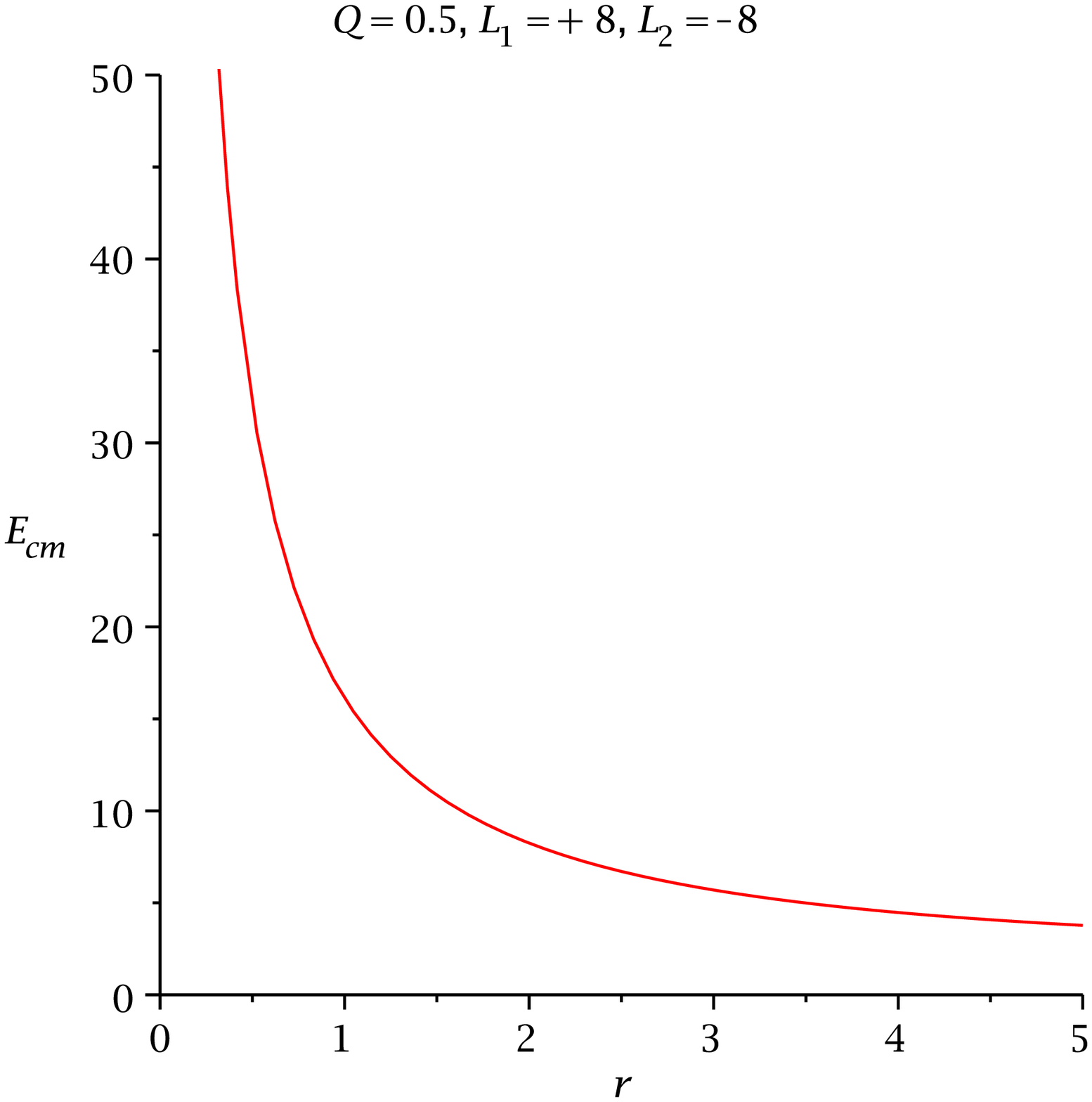}}\qquad
\subfigure[][]{\includegraphics[width=0.45\textwidth]{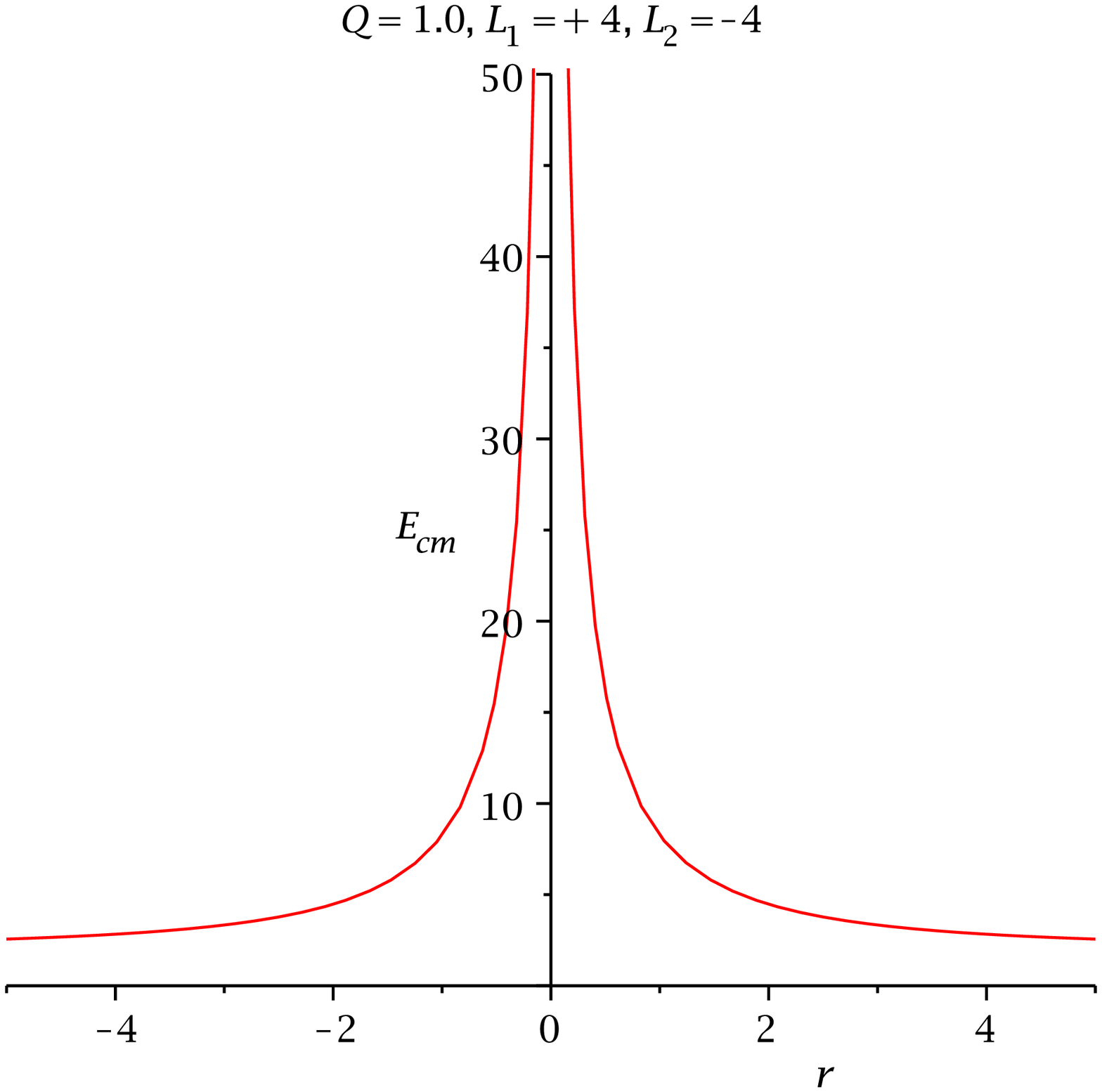}}
\end{center}
\caption{The figure shows the variation  of ${\cal E}_{cm}$  with $r$ for RN BH
with $Q=0.5$ and $L_{1}=8, L_{2}=-8$, $Q=1.0$ and $L_{1}=4, L_{2}=-4$
 \label{rnecm}}
\end{figure}

\begin{figure}[t]
\begin{center}
\subfigure[][]{\includegraphics[width=0.45\textwidth]{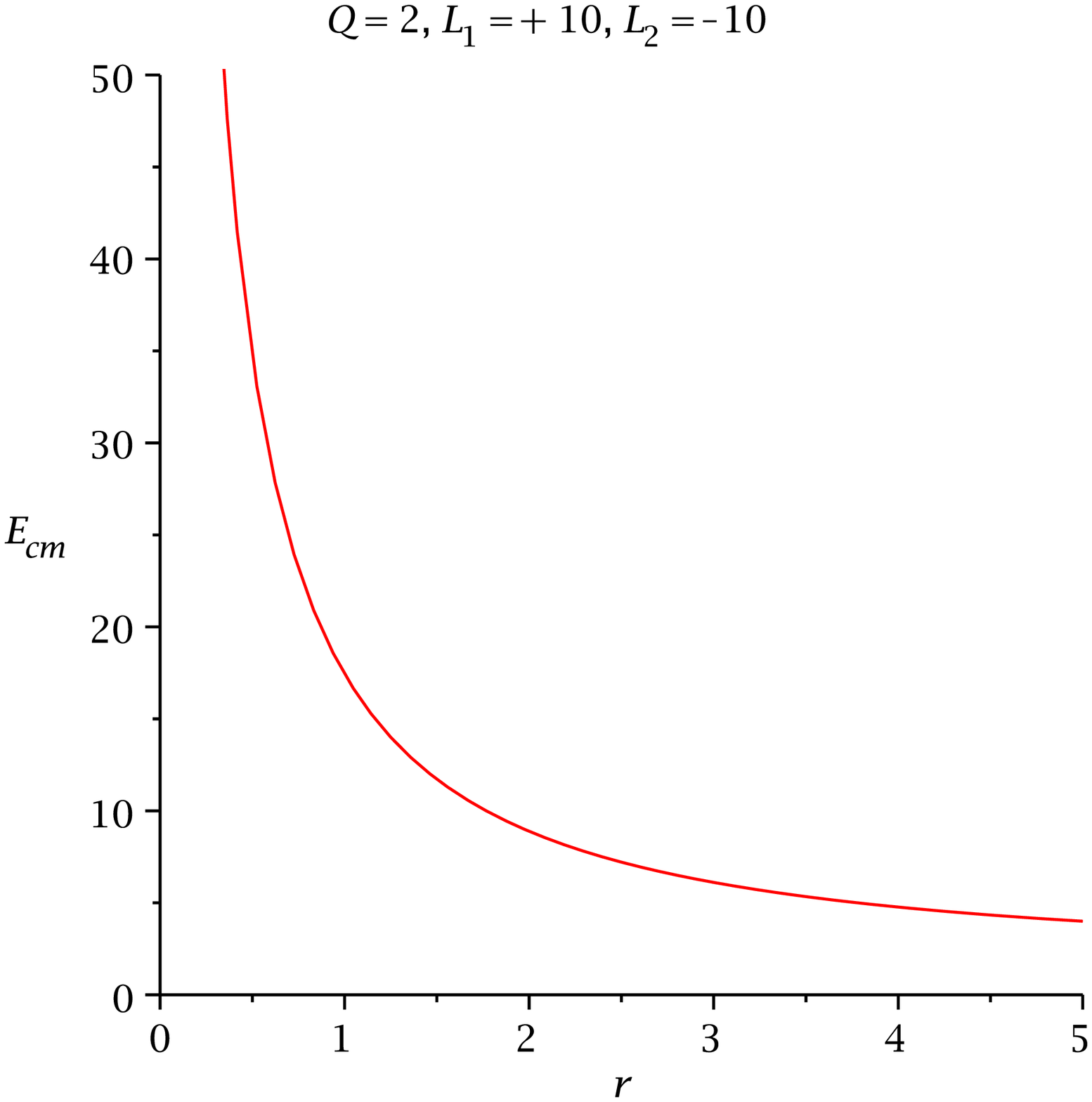}}
\end{center}
\caption{The figure shows the variation  of ${\cal E}_{cm}$  with $r$
for RN BH with $Q=2.0$ and $L_{1}=10, L_{2}=-10$.
 \label{rnecm1}}
\end{figure}

As we have assumed ${\cal E}_{1}={\cal E}_{2}=1$ previously, and substituting  the value of ${\cal G}(r)$ one could  obtain finally the CM energy near the  event horizon($r_{+}$) for non-extremal RN space-time(See Figs. 3-4)
\begin{eqnarray}
{\cal E}_{cm}\mid_{r\rightarrow r_{+}} &=& \sqrt{2}m_{0}\sqrt{\frac{4r_{+}^2+(L_{1}-L_{2})^{2}}{2r_{+}^2}} ~.\label{cm6}
\end{eqnarray}

Again we have already known from \cite{bsw}, the maximum CM energy strictly depend upon the values of critical angular momentum such that the particles can reach the event horizon with maximum tangential velocity. Thus the critical angular momentum and the critical radius could
be calculated from the radial effective potential. Now we impose the following criterion for determining the critical values of angular momentum i.e. :
\begin{eqnarray}
 \dot{r}^2r^4 &=& ({\cal E}^2-1)r^4+2Mr^3-(L^2+Q^2)r^2+2M L^2r-Q^2L^2 = 0
\end{eqnarray}
Now setting ${\cal E}^2=1$ for marginal case, the equation turns out to be
\begin{eqnarray}
 2Mr^3-(L^2+Q^2)r^2+2M L^2r-Q^2 L^2 &=& 0   ~.\label{angnrn}
\end{eqnarray}
For non-extremal RN BH, the critical values of angular momentum can be determined by finding the real root of the above equation and using the following condition:
\begin{eqnarray}
 (M^2-Q^2)L^6-(3Q^4-20M^2Q^2+16M^4)L^4-(8M^2+3Q^2)Q^4 L^2-Q^8 &=& 0
\end{eqnarray}
It can be easily seen that in the limit $Q=0$, we retrieve the critical values of angular momentum for Schwarzschild BH:
\begin{eqnarray}
 L &=& \pm 4M
\end{eqnarray}
Since, we are interested in this work for extremal RN BH, thus it can be  easily calculated the critical values of angular momentum parameter in the extremal limit $M=Q$. Therefore the range of $L$ for in-falling geodesics is
\begin{eqnarray}
-\sqrt{\frac{11+5\sqrt{5}}{2}}M \leq L \leq \sqrt{\frac{11+5\sqrt{5}}{2}}M
\end{eqnarray}

For extremal RN space-time, the values of CM energy near the horizon is given by
\begin{eqnarray}
{\cal E}_{cm}\mid_{r\rightarrow M} &=& \sqrt{2}m_{0}\sqrt{\frac{4M^2+(L_{1}-L_{2})^{2}}{2M^2}} ~.\label{cm8}
\end{eqnarray}

Since here we have assumed that the collision point is at $r=M$ and two particles have the angular momentum which is given by the following equation:
\begin{eqnarray}
L &=& \pm \sqrt{\frac{r^2(2Mr-Q^2)}{r^2-2Mr+Q^2}}
\end{eqnarray}
In the extremal limit it goes to
\begin{eqnarray}
L &=& \pm \sqrt{\frac{Mr^2(2r-M)}{(r-M)^2}}
\end{eqnarray}
It was already mentioned in \cite{bsw}, a new phenomenon would appear if one of the
particles participating in the collision has the critical angular momentum. If one
of the particles have the diverging angular momentum at the horizon i.e.

\begin{eqnarray}
 L_{1} \mid_{r=M} &=&  \sqrt{\frac{Mr^2(2r-M)}{(r-M)^2}} \rightarrow \infty
\end{eqnarray}
then the corresponding value of the CM energy (using Eq. {\ref{cm8}}) for extremal RN BH
at the extremal horizon is
\begin{eqnarray}
{\cal E}_{cm}\mid_{r\rightarrow M}
&=& \sqrt{2}m_{0}\sqrt{\frac{4M^2+(L_{1}-L_{2})^{2}}{2M^2}} \rightarrow \infty
\end{eqnarray}
Thus for the extremal RN BH, we get the infinite amounts of CM energy.

Now we will see what happens the CM energy if we choose the collision point
to be ISCO.
It is well known that for extremal RN BH the ISCO is at $r=4M$ \cite{ms}.
Now if we consider the collision point to be ISCO then the corresponding value of
CM energy at the ISCO for extremal RN space-time is given by
\begin{eqnarray}
{\cal E}_{cm}\mid_{r\rightarrow 4M} &=&
\sqrt{2}m_{0}\sqrt{\frac{400M^2-9L_{1}L_{2}-\sqrt{112M^2-9L_{1}^2}\sqrt{112M^2-9L_{2}^2}}
{144M^2}} ~.\label{cm9}
\end{eqnarray}

For the critical values of angular momentum $L_{1}=\sqrt{\frac{11+5\sqrt{5}}{2}}M$ and
$L_{2}=-\sqrt{\frac{11+5\sqrt{5}}{2}}M$, the CM energy is found to be

\begin{eqnarray}
{\cal E}_{cm}\mid_{r\rightarrow 4M} &=& \sqrt{2}m_{0} \sqrt{\frac{387+45\sqrt{5}}{144}}=2.60m_{0}
\end{eqnarray}

Similarly, one may compute the MBCO for RN BH by setting ${\cal E}_{0}^2=1$ which turns out to be $r_{mb}=\left(\frac{3+\sqrt{5}}{2}\right)M$. Thus when the collision point to be MBCO, the CM energy is found to be at $r=r_{mb}$:

\begin{eqnarray}
{\cal E}_{cm}\mid_{r\rightarrow r_{mb}} &=&
\sqrt{2}m_{0}\sqrt{\frac{(10+4\sqrt{5})(7+3\sqrt{5})M^2-(6+2\sqrt{5})L_{1}L_{2}-
\Lambda_{1}\Lambda_{2}}
{(3+\sqrt{5})(7+3\sqrt{5})M^2}}
~.\label{cm10}\\
\mbox{where} \nonumber \\
\Lambda_{1} &=& \sqrt{(58+26\sqrt{5})M^2-(6+2\sqrt{5})L_{1}^2} \\
\Lambda_{2} &=& \sqrt{(58+26\sqrt{5})M^2-(6+2\sqrt{5})L_{2}^2}
\end{eqnarray}
For the critical values of angular momentum $L_{1}=\sqrt{\frac{11+5\sqrt{5}}{2}}M$ and
$L_{2}=-\sqrt{\frac{11+5\sqrt{5}}{2}}M$, the CM energy is at $r_{mb}$ found to be
\begin{eqnarray}
{\cal E}_{cm}\mid_{r=r_{mb}} &=& \sqrt{2}m_{0} \sqrt{\frac{47+21\sqrt{5}}{9+4\sqrt{5}}}=3.23m_{0}
\end{eqnarray}

Now we can compare the results obtained as
\begin{eqnarray}
{\cal E}_{cm}\mid_{r_{+}=M}: {\cal E}_{cm}\mid_{r_{mb}=\left(\frac{3+\sqrt{5}}{2}\right)M} : {\cal E}_{cm}\mid_{r_{ISCO}=4M}
&=& \infty : 3.23 : 2.6
\end{eqnarray}
Thus we get,
\begin{eqnarray}
{\cal E}_{cm}\mid_{r_{+}}>{\cal E}_{cm}\mid_{r_{mb}}>{\cal E}_{cm}\mid_{r_{ISCO}}
\end{eqnarray}
It is clearly evident that CM energy is diverging at the event horizon and finite
at the MBCO and at ISCO.

In the limit $L_{1}=L_{2}=0$, we obtain the following equality:
\begin{eqnarray}
{\cal E}_{cm}\mid_{r_{+}}={\cal E}_{cm}\mid_{r_{mb}}={\cal E}_{cm}\mid_{r_{ISCO}}=2m_{0}
\end{eqnarray}

\section{CM Energy of Particle Collision near the ISCO of a Schwarzschild BH:}

Here, we shall extend our analysis for Schwarzschild BH and calculate the CM energy
of the colliding particles at various collision points. First we choose the collision
point to be horizon, then we choose the collision point to be ISCO and finally we
have chosen the collision point is at MBCO.
To proceeds, we first write the metric  of the Schwarzschild BH in Schwarzschild
coordinates are
\begin{eqnarray}
ds^2 &=& -\left(1-\frac{2M}{r}\right)dt^{2}+\left(1-\frac{2M}{r}\right)^{-1}dr^{2}
+r^2\left(d\theta^{2}+\sin^{2}\theta d\phi^{2}\right) ~.\label{scg}
\end{eqnarray}
The BH has an event horizon is located at $r=2M$.
For this BH, the equatorial time-like geodesics are described by
\begin{eqnarray}
  u^{t} &=& \frac{{{\cal E}}}{1-\frac{2M}{r}}  \\
  u^{r} &=& \pm \sqrt{{\cal E}^{2}-\left(1-\frac{2M}{r}\right)
   \left(1+\frac{L^{2}}{r^2}\right)} \\
  u^{\theta} &=& 0 \\
  u^{\phi} &=& \frac{L}{r^2} ~.\label{utur4}
\end{eqnarray}
and the components of four velocity are(See Fig. 5a)

\begin{eqnarray}
u^{\mu}_{1} &=& \left( \frac{{\cal E}_{1}}{1-\frac{2M}{r}},~ -Z_{1},~ 0,~\frac{L_{1}}{r^{2}}\right)
~.\label{u31}\\
u^{\mu}_{2} &=& \left( \frac{{\cal E}_{2}}{1-\frac{2M}{r}},~ -Z_{2},~ 0,~\frac{L_{2}}{r^{2}}\right)
~.\label{u5}
\end{eqnarray}







\begin{figure}[t]
\begin{center}
\subfigure[][]{\includegraphics[width=0.45\textwidth]{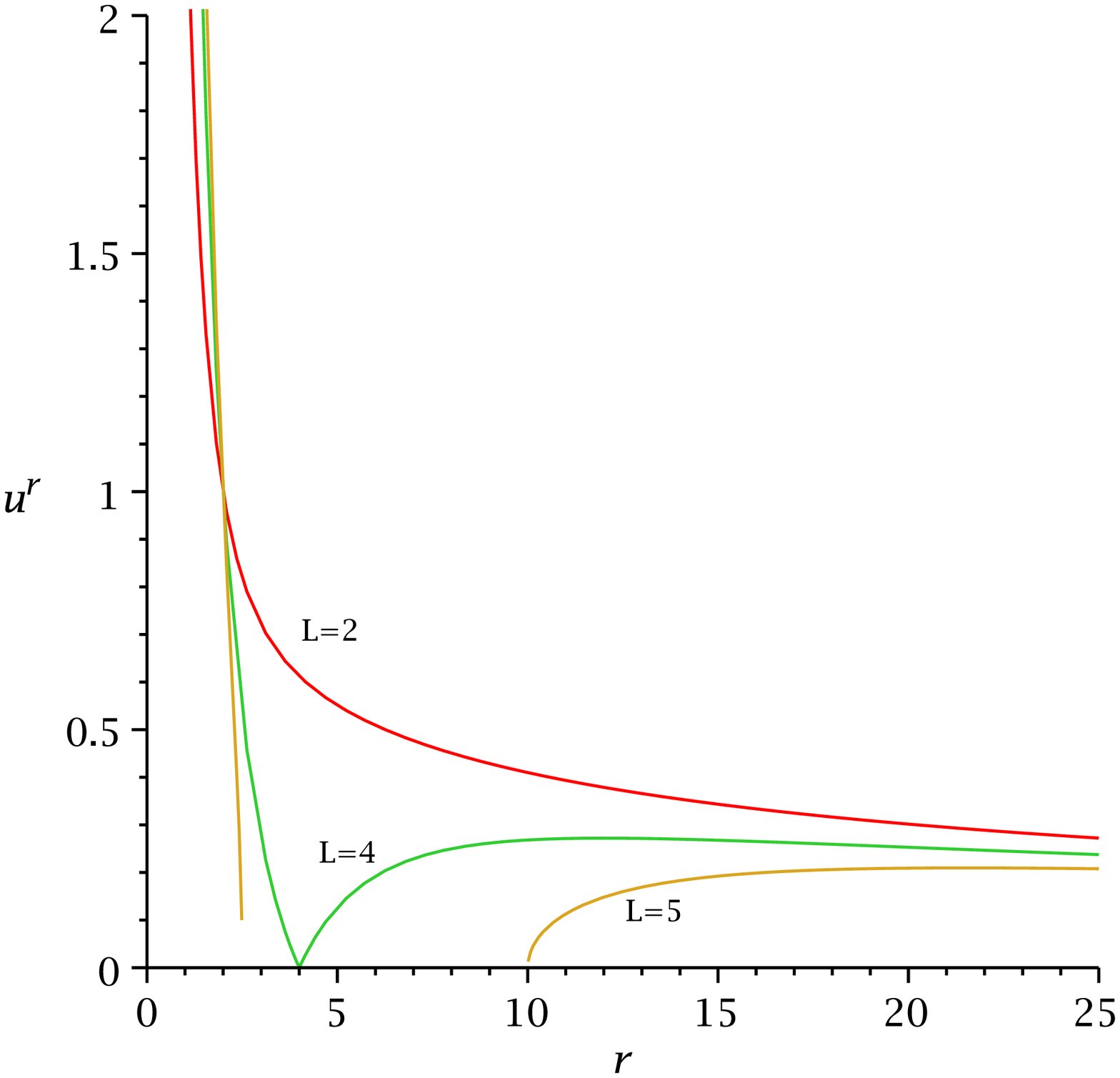}}\qquad
\subfigure[][]{\includegraphics[width=0.45\textwidth]{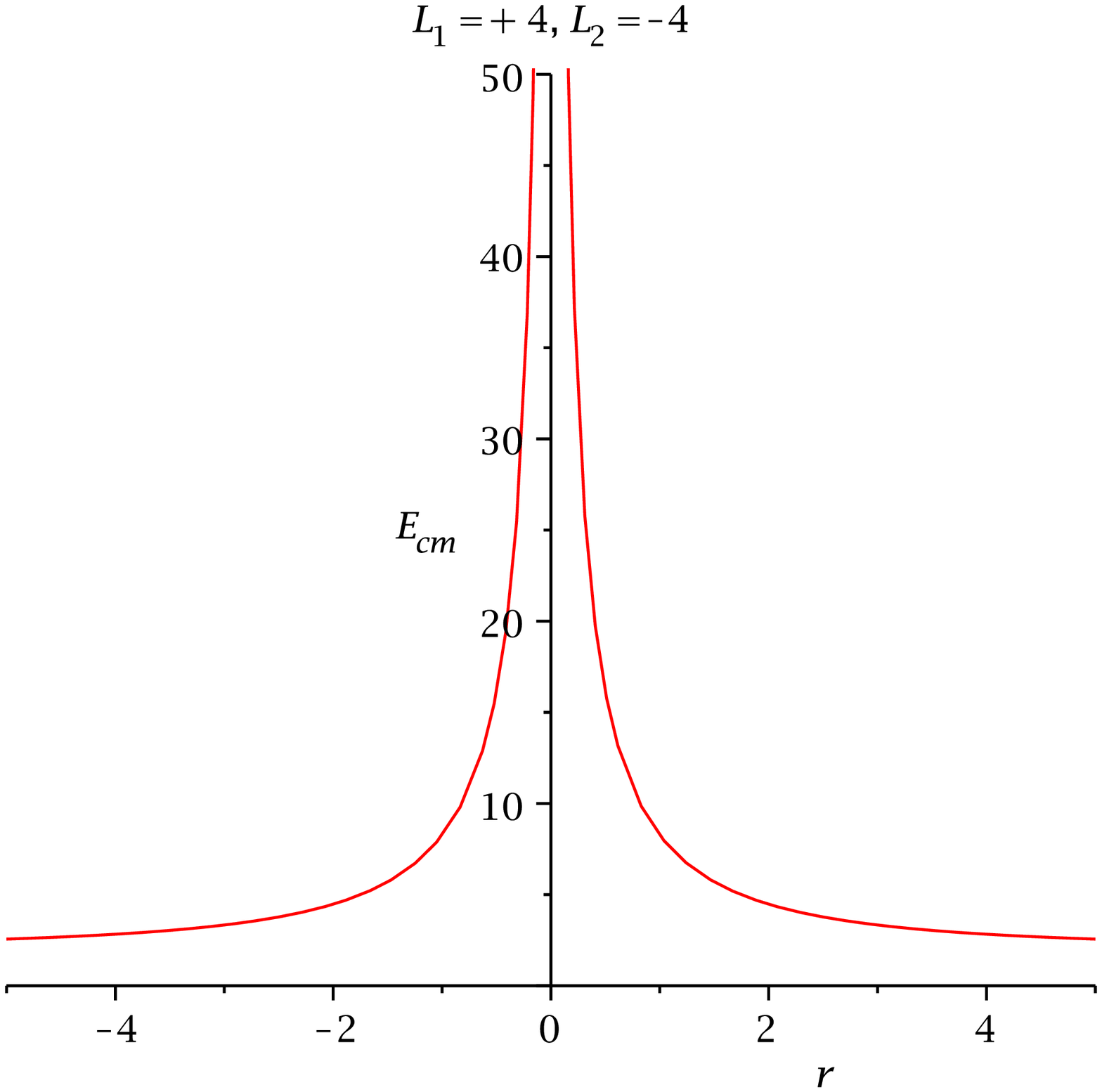}}
\end{center}
\caption{The LHS figure shows the variation  of $\dot{r}$  with $r$
for Schwarzschild BH and The RHS figure shows the variation  of ${\cal E}_{cm}$  with $r$
for Schwarzschild BH.
 \label{schcase}}
\end{figure}

\begin{figure}[t]
\begin{center}
\subfigure[][]{\includegraphics[width=0.45\textwidth]{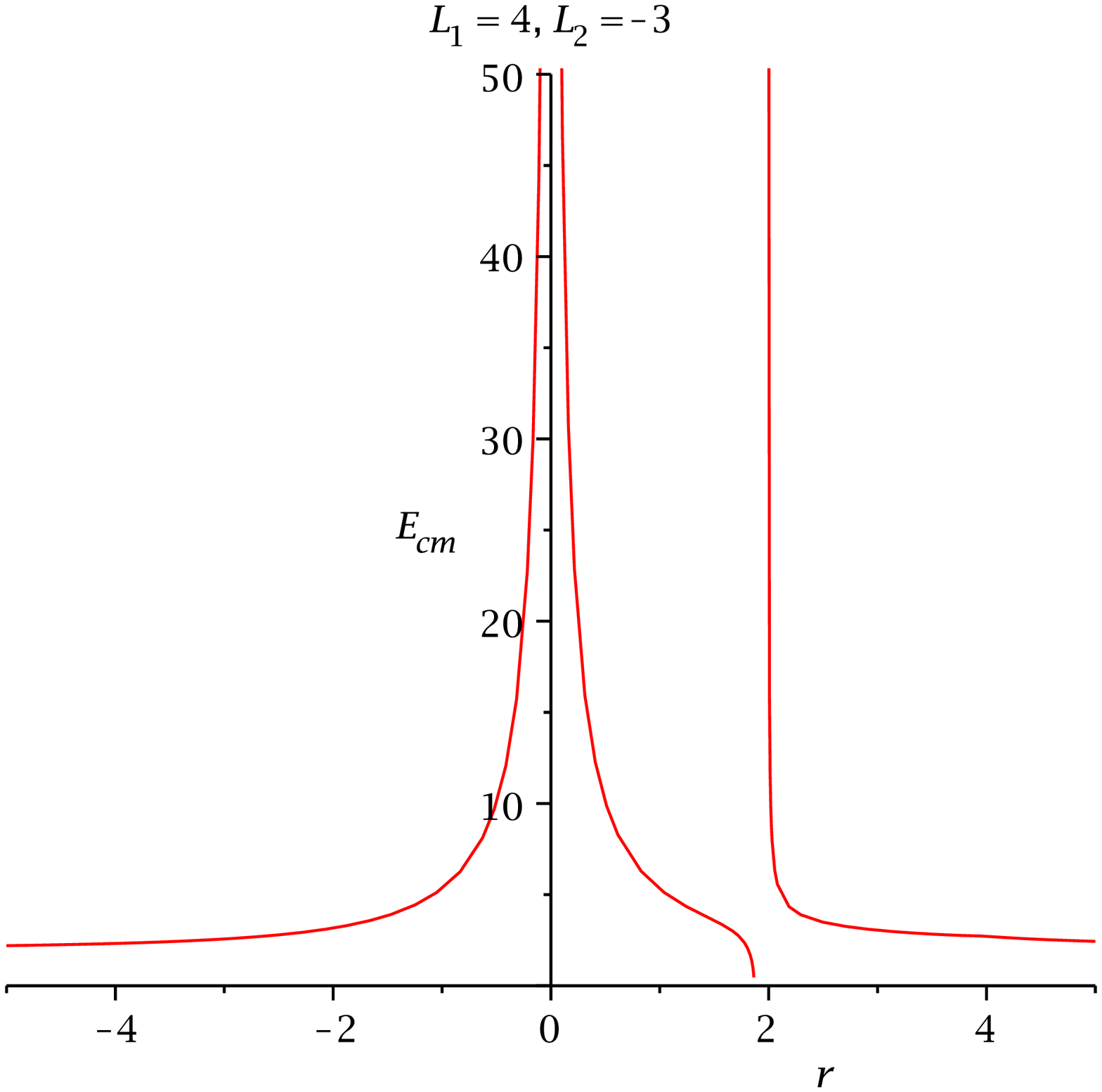}}\qquad
\subfigure[][]{\includegraphics[width=0.45\textwidth]{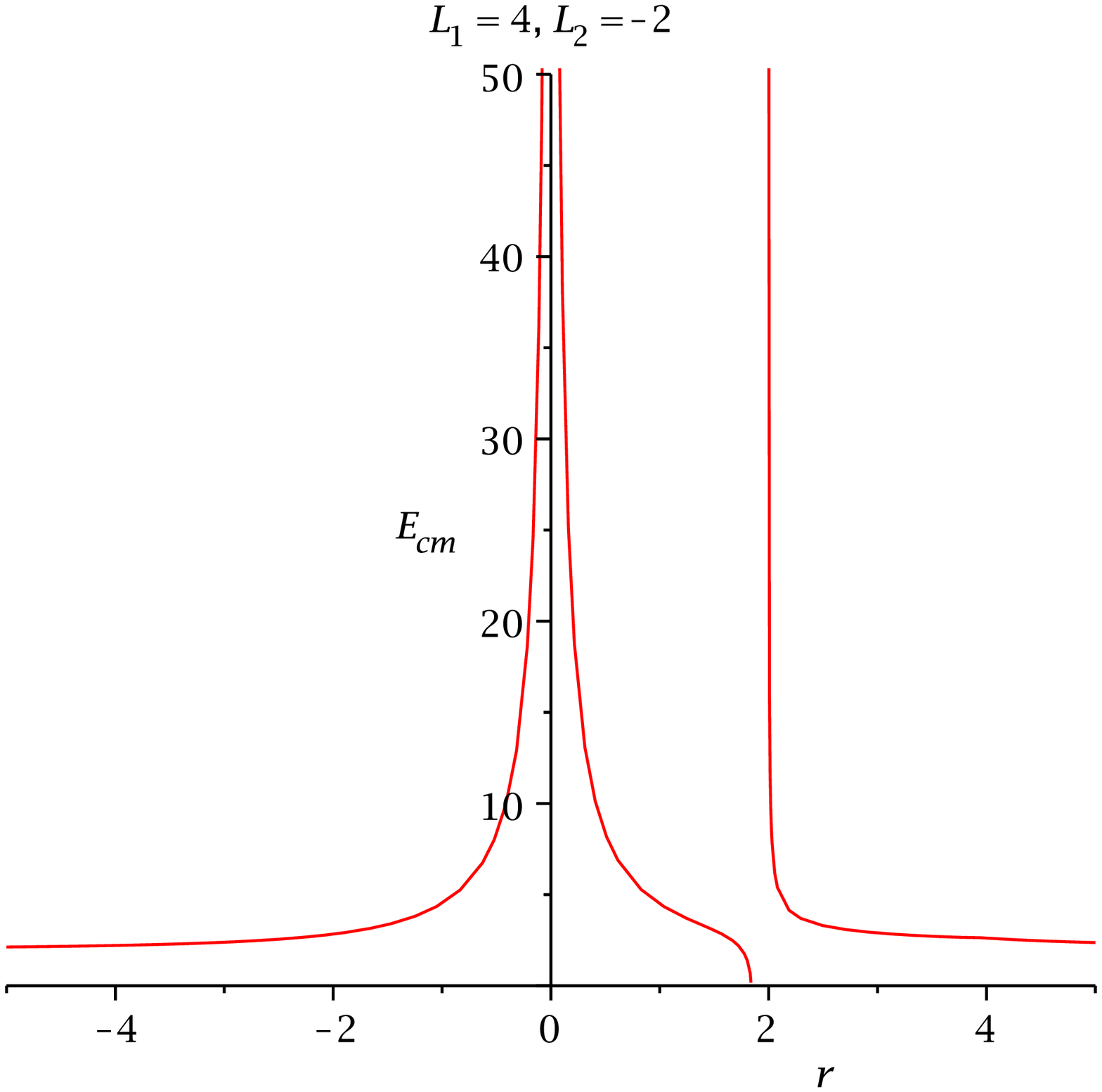}}
\end{center}
\caption{The figures show the variation  of ${\cal E}_{cm}$  with $r$
for Schwarzschild BH.
 \label{schcase1}}
\end{figure}

\begin{figure}[t]
\begin{center}
\subfigure[][]{\includegraphics[width=0.45\textwidth]{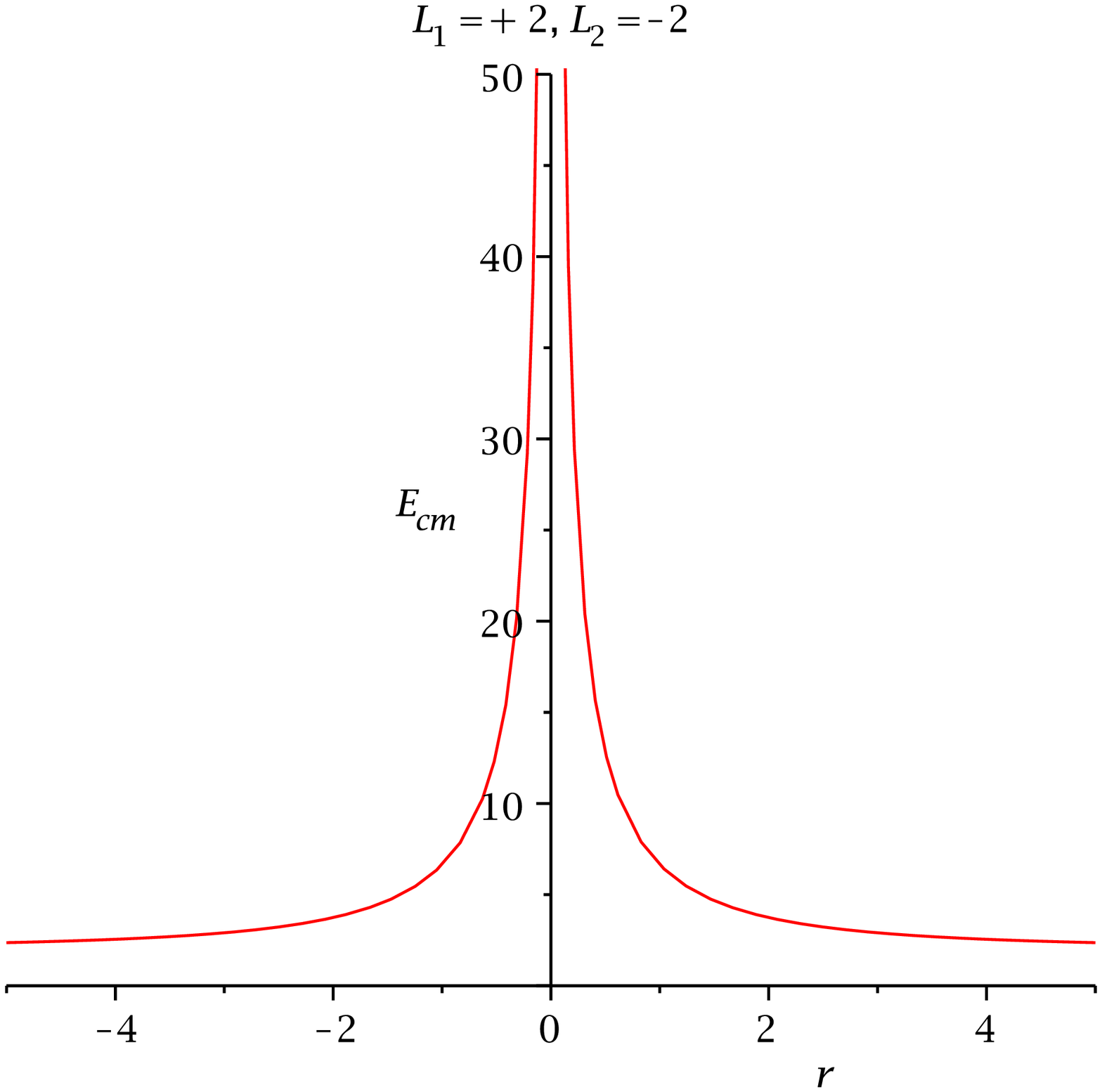}}
\end{center}
\caption{The figure shows the variation  of ${\cal E}_{cm}$  with $r$
for Schwarzschild BH.
 \label{schcase2}}
\end{figure}

Since the Schwarzschild space-time is the special case of RN space-time, therefore one can compute the CM energy(See Figs. 5b-7) using the equation (\ref{cm4}) as
\begin{eqnarray}
\left(\frac{{\cal E}_{cm}}{\sqrt{2}m_{0}}\right)^{2} &=&  1 +\frac{{\cal E}_{1}{\cal E}_{2}}{\left(1-\frac{2M}{r}\right)}-
\frac{Z_{1}Z_{2}}{\left(1-\frac{2M}{r}\right)} -\frac{L_{1}L_{2}}{r^2}~.\label{cm10.5}\\
\mbox{where}\\\nonumber
Z_{1} &=& \sqrt{{\cal E}_{1}^{2}-\left(1-\frac{2M}{r}\right)\left(1+\frac{L_{1}^{2}}{r^2}\right)}\\
Z_{2} &=& \sqrt{{\cal E}_{2}^{2}-\left(1-\frac{2M}{r}\right)\left(1+\frac{L_{2}^{2}}{r^2}\right)}
\end{eqnarray}

Setting ${\cal E}_{1}={\cal E}_{2}=1$ as is, one may obtain the CM energy near the event
horizon  for Schwarzschild BH (first calculated in \cite{bsw}) is given by
\begin{eqnarray}
{\cal E}_{cm}\mid_{r\rightarrow 2M} &=& \sqrt{2}m_{0}\sqrt{\frac{16M^2+(L_{1}-L_{2})^{2}}{8M^2}} ~.\label{cmSch}
\end{eqnarray}
When $L_{1}=4M$ and $L_{2}=-4M$, the maximum CM energy for
Schwarzschild BH is found to be ${\cal E}_{cm}\mid_{r\rightarrow 2M}=2\sqrt{5}m_{0}$.

Again, we know the ISCO for Schwarzschild BH is  $r=6M$, thus the CM energy
at ISCO is calculated to be
\begin{eqnarray}
{\cal E}_{cm}\mid_{r\rightarrow 6M} &=&  \sqrt{2}m_{0}\sqrt{\frac{(90M^2-L_{1}L_{2})-\sqrt{18M^2-L_{1}^2}\sqrt{18M^2-L_{2}^2}}
{36M^2}} ~.\label{cm11}
\end{eqnarray}
For the critical values of angular momentum $L_{1}=4M$ and $L_{2}=-4M$, the CM
energy is given by
\begin{eqnarray}
{\cal E}_{cm}\mid_{r\rightarrow r_{ISCO}=6M} &=&\sqrt{\frac{52}{9}}m_{0} ~.\label{cm12}
\end{eqnarray}

It is well known that the MBCO  for Schwarzschild BH is occur at $r_{mb}=4M$, thus
the CM energy is found to be:
\begin{eqnarray}
{\cal E}_{cm}\mid_{r\rightarrow 4M} &=&  \sqrt{2}m_{0}\sqrt{\frac{(48M^2-L_{1}L_{2})-\sqrt{16M^2-L_{1}^2}\sqrt{16M^2-L_{2}^2}}
{16M^2}} ~.\label{mbcm}
\end{eqnarray}
If we take $L_{1}=L_{2}=0$, the CM energy is found to be ${\cal E}_{cm}=2m_{0}$. Whenever
we take $L_{1}=4M$ and $L_{2}=-4M$, we have the CM energy
\begin{eqnarray}
{\cal E}_{cm}\mid_{r\rightarrow r_{mb}=4M} &=& 2\sqrt{2}m_{0} ~.\label{cm13}
\end{eqnarray}
Now we may compare the results obtained  as
\begin{eqnarray}
{\cal E}_{cm}\mid_{r_{+}=2M}:  {\cal E}_{cm}\mid_{r_{mb}=4M} : {\cal E}_{cm}\mid_{r_{ISCO}=6M}
&=& 2\sqrt{5} : 2\sqrt{2} : \sqrt{\frac{52}{9}} \\
&=& \sqrt{5} : \sqrt{2} : \frac{\sqrt{13}}{3} \\
&=& 2.23 : 1.41 : 1.20
\end{eqnarray}
Thus we have found,
\begin{eqnarray}
{\cal E}_{cm}\mid_{r_{+}}>{\cal E}_{cm}\mid_{r_{mb}}> {\cal E}_{cm}\mid_{r_{ISCO}}
\end{eqnarray}
It is clearly evident that CM energy is maximum at the event horizon than the
MBCO and ISCO.

\section{CM energy of Particle collision near the horizon of the GMGHS BH:}

In an earlier paper \cite{ppstring}, we showed that a string BHs may act
as a particle accelerators to arbitrarily high center-of-mass energy. Also, we
proved that for extremal GMGHS space-time the center of mass energy of collision
at $r \equiv r_{ISCO}=r_{ph}=r_{mb}=r_{hor}=2M$ is arbitrarily large. Here, we
would like to discuss the same more elaborately and graphically which was not
given in the previous work.

To proceed it, first we consider the time-like geodesics(See Figs.8-9) on the
equatorial plane \cite{ppstring} reads
\begin{eqnarray}
  u^{t} &=& \frac{{{\cal E}}}{{\cal H}(r)}  \\
  u^{r} &=& \pm \sqrt{{\cal E}^{2}-{\cal H}(r)\left(1+\frac{L^{2}}{r(r-b)}\right)} \label{eff1}\\
  u^{\theta} &=& 0 \\
  u^{\phi} &=& \frac{L}{r(r-b)} ~.\label{uturuphi}
\end{eqnarray}
where ${\cal H}(r)=1-\frac{2M}{r}$ and $b=\frac{Q^{2}}{M}e^{-2\phi_{0}}$.
and
\begin{eqnarray}
u^{\mu}_{1}= \left(\frac{{\cal E}_{1}}{{\cal H}(r)},~ -X_{1},~ 0,~\frac{L_{1}}{r^{2}}\right) ~.\label{u11}\\
u^{\mu}_{2}= \left(\frac{{\cal E}_{2}}{{\cal H}(r)},~ -X_{2},~ 0,~\frac{L_{2}}{r^{2}}\right) ~.\label{u22}
\end{eqnarray}





\begin{figure}[t]
\begin{center}
\subfigure[][]{\includegraphics[width=0.45\textwidth]{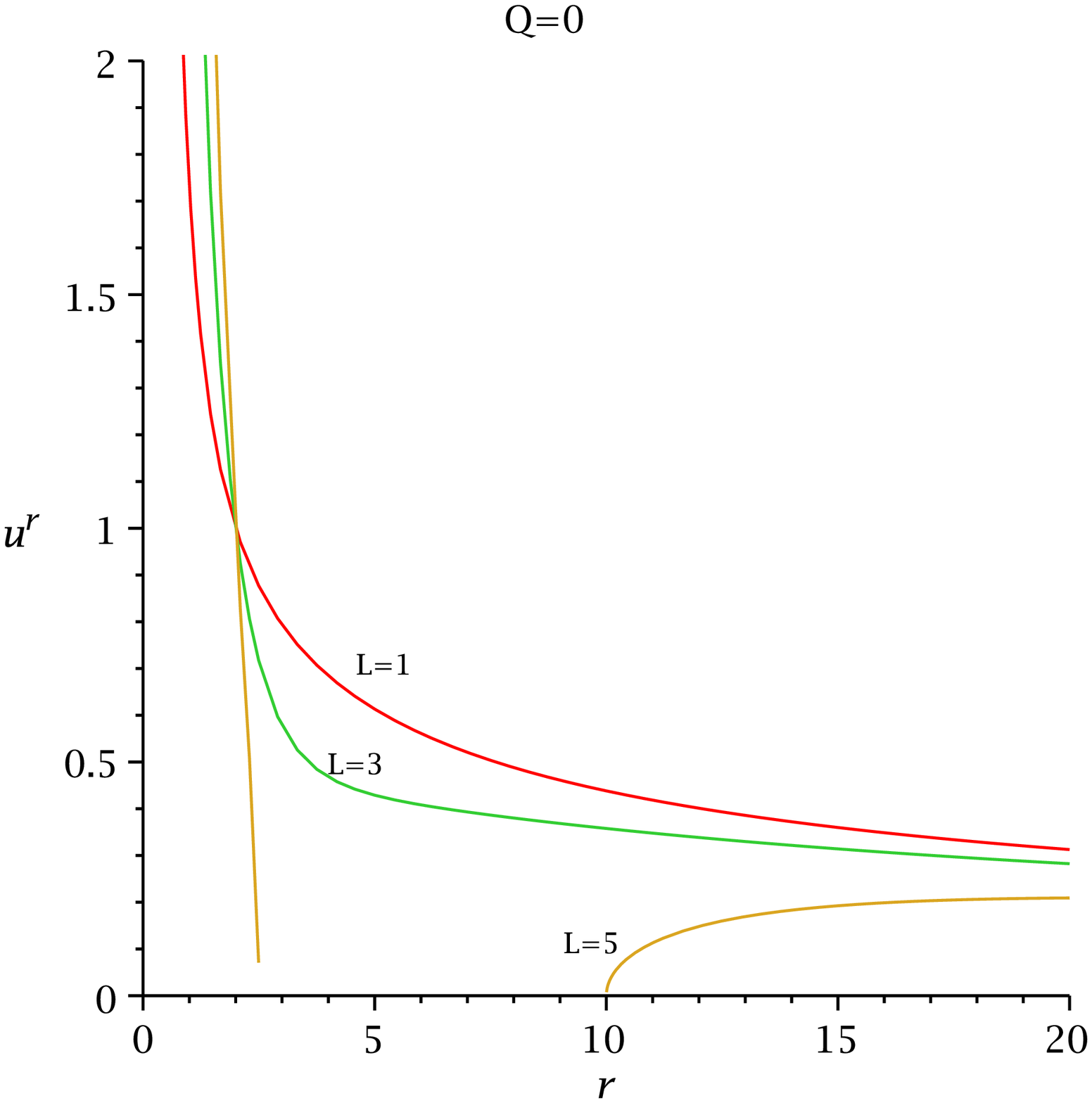}}\qquad
\subfigure[][]{\includegraphics[width=0.45\textwidth]{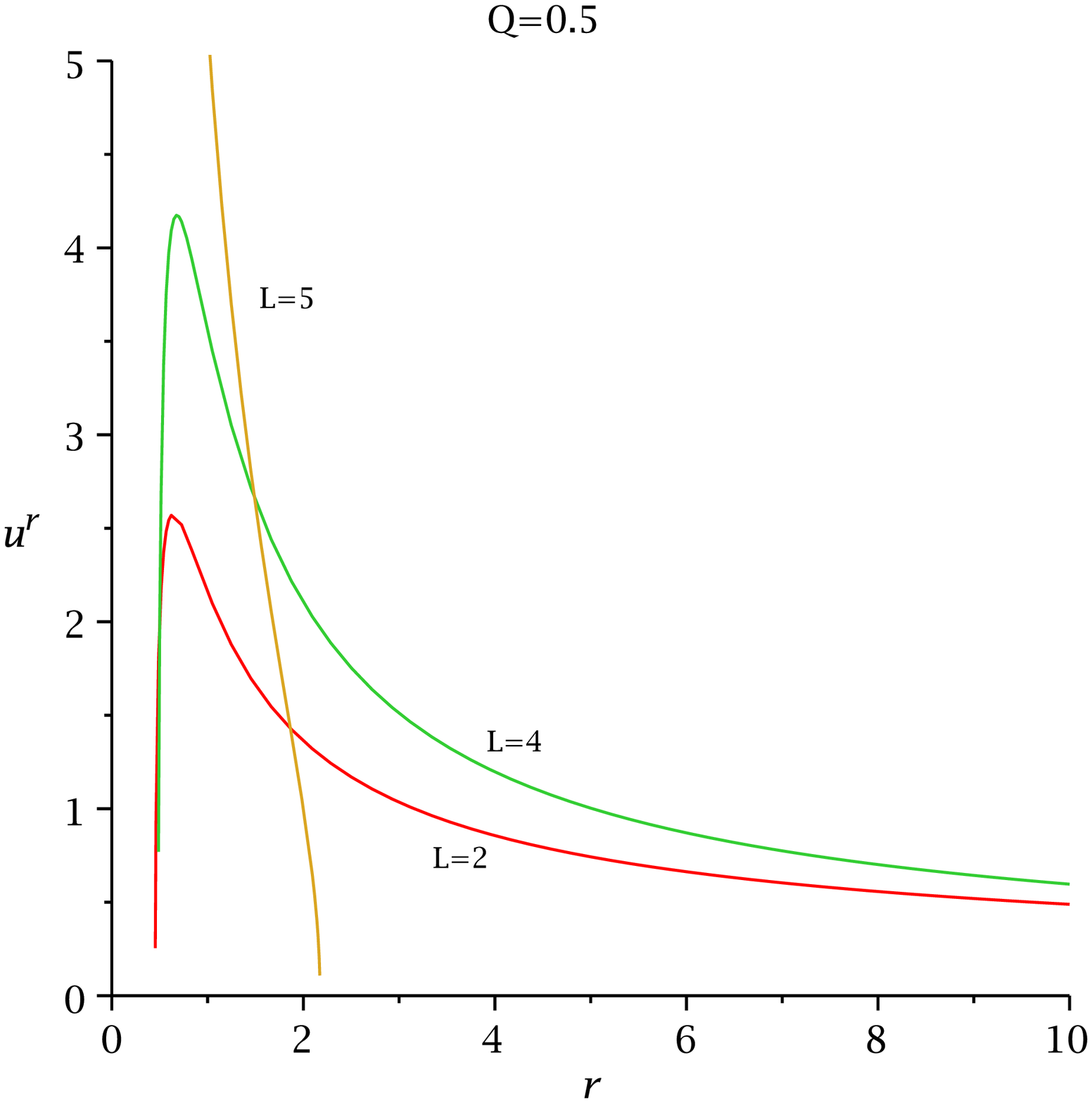}}
\end{center}
\caption{The figure shows the variation  of $\dot{r}$  with $r$
for GMGHS BH.
 \label{gmgcase}}
\end{figure}

\begin{figure}[t]
\begin{center}
\subfigure[][]{\includegraphics[width=0.45\textwidth]{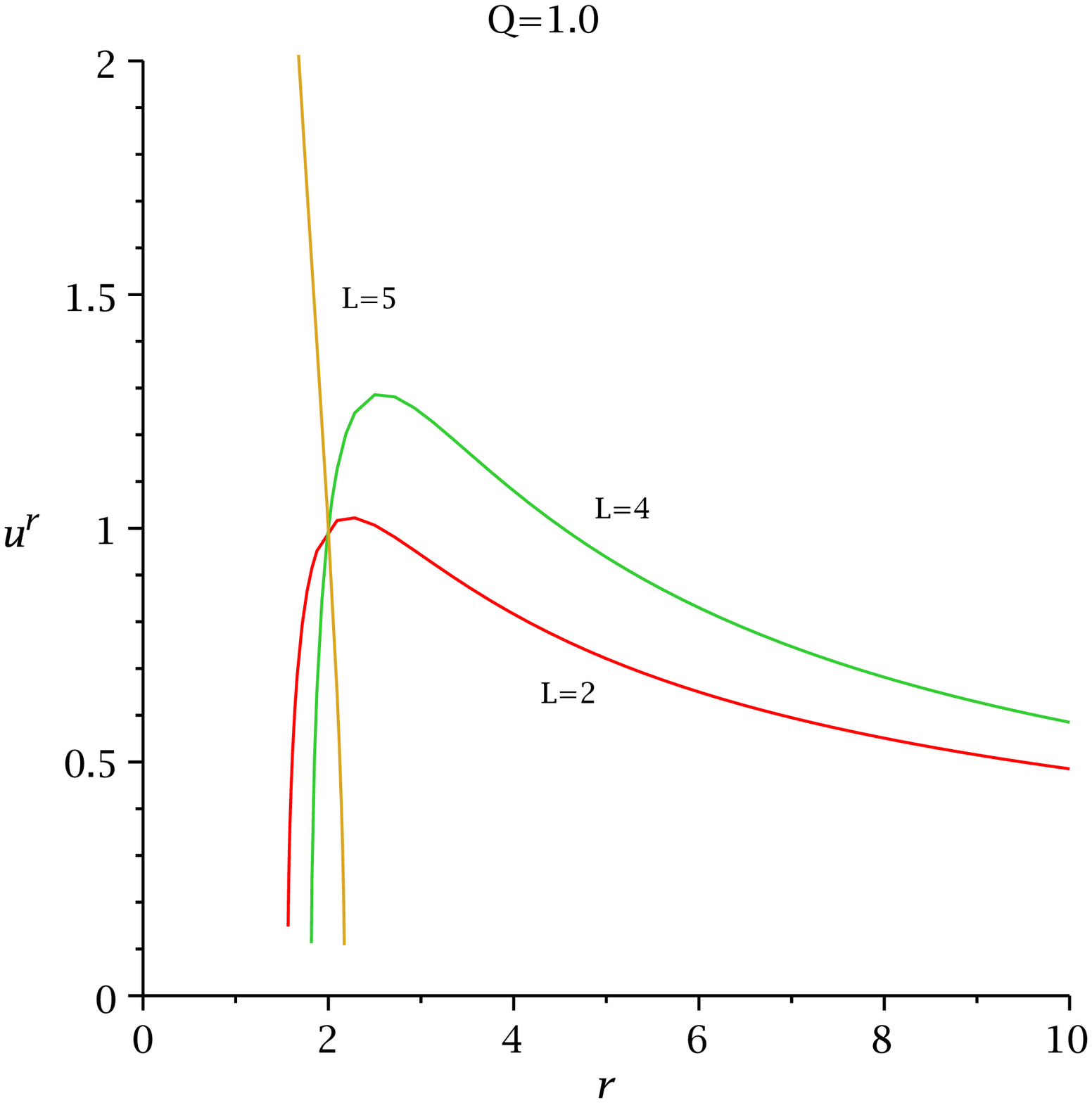}}\qquad
\subfigure[][]{\includegraphics[width=0.45\textwidth]{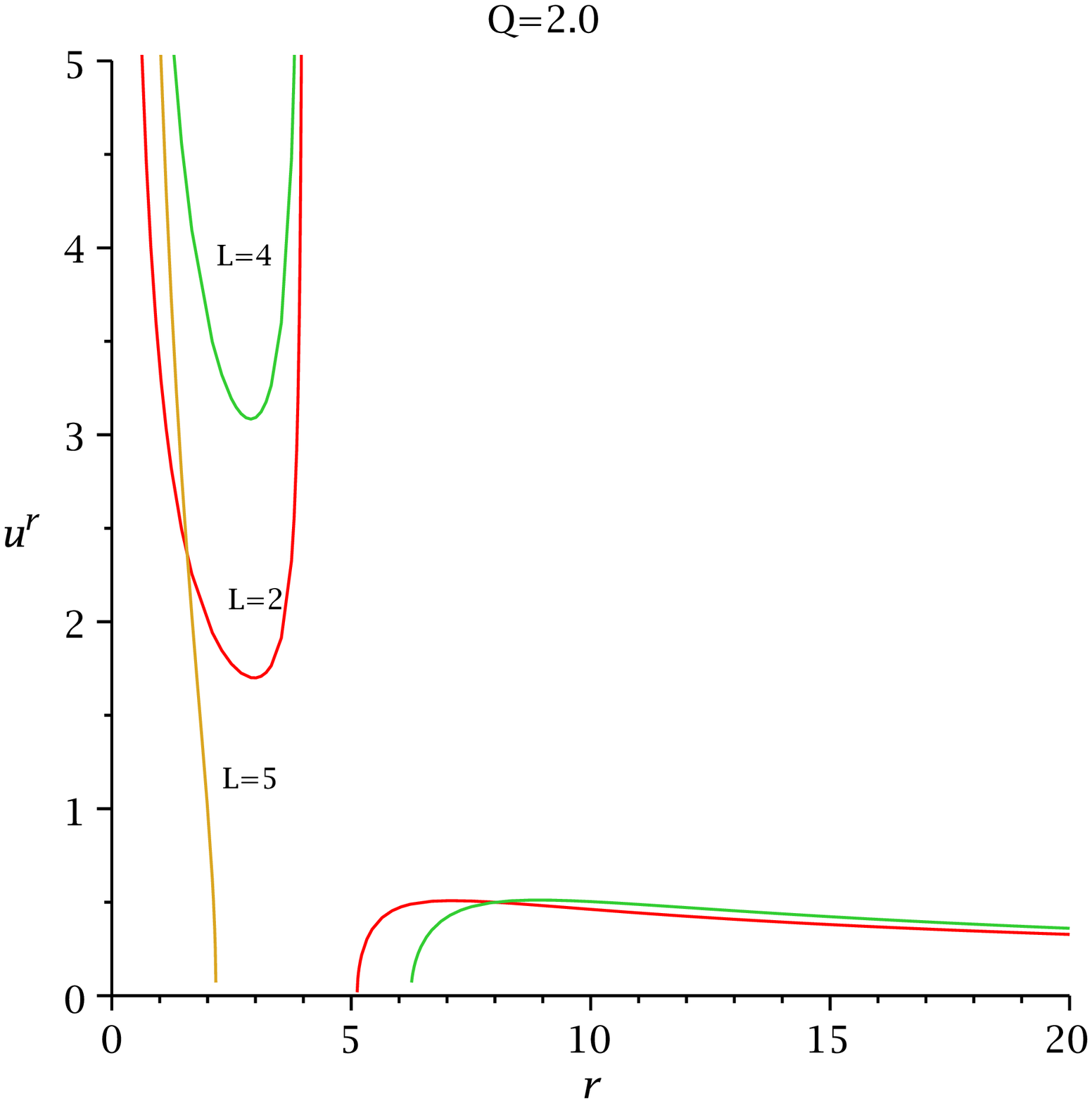}}
\end{center}
\caption{The figure shows the variation  of $\dot{r}$  with $r$
for GMGHS BH.
 \label{gmgcase1}}
\end{figure}

Therefore using(\ref{cm}), we obtain the CM energy(See Figs. 10-11) for this collision:
\begin{eqnarray}
\left(\frac{{\cal E}_{cm}}{\sqrt{2}m_{0}}\right)^{2} &=&  1 +\frac{{\cal E}_{1}{\cal E}_{2}}{{\cal H}(r)}
-\frac{X_{1}X_{2}}{{\cal H}(r)}-\frac{L_{1}L_{2}}{r(r-b)} ~.\label{cm14} \\
\mbox{where}\\\nonumber
X_{1} &=& \sqrt{{\cal E}_{1}^{2}-{\cal H}(r)\left(1+\frac{L_{1}^{2}}{r(r-b)}\right)} \\
X_{2} &=& \sqrt{{\cal E}_{2}^{2}-{\cal H}(r)\left(1+\frac{L_{2}^{2}}{r(r-b)}\right)}
\end{eqnarray}
As we have assumed ${\cal E}_{1}={\cal E}_{2}=1$ previously and substituting ${\cal H}(r)=1-\frac{2M}{r}$,  we obtain finally the CM energy near the horizon:
\begin{eqnarray}
{\cal E}_{cm} = \sqrt{2}m_{0}\sqrt{\frac{8M(2M-b)+(L_{1}-L_{2})^{2}}{4M(2M-b)}} ~.\label{cm15}
\end{eqnarray}
Whenever we taking the extremal limit $b=2M$, we get the CM energy near the horizon:
\begin{eqnarray}
{\cal E}_{cm} &=&  \sqrt{2}m_{0}\sqrt{\frac{8M(2M-b)+(L_{1}-L_{2})^{2}}{4M(2M-b)}} \\
{\cal E}_{cm} &\longmapsto &  \infty \nonumber\\
\end{eqnarray}
which implies that the CM energy of collision  for extremal dilation black hole blows up as we approach the extremal limit.




\newpage
\begin{figure}[t]
\begin{center}
\subfigure[][]{\includegraphics[width=0.45\textwidth]{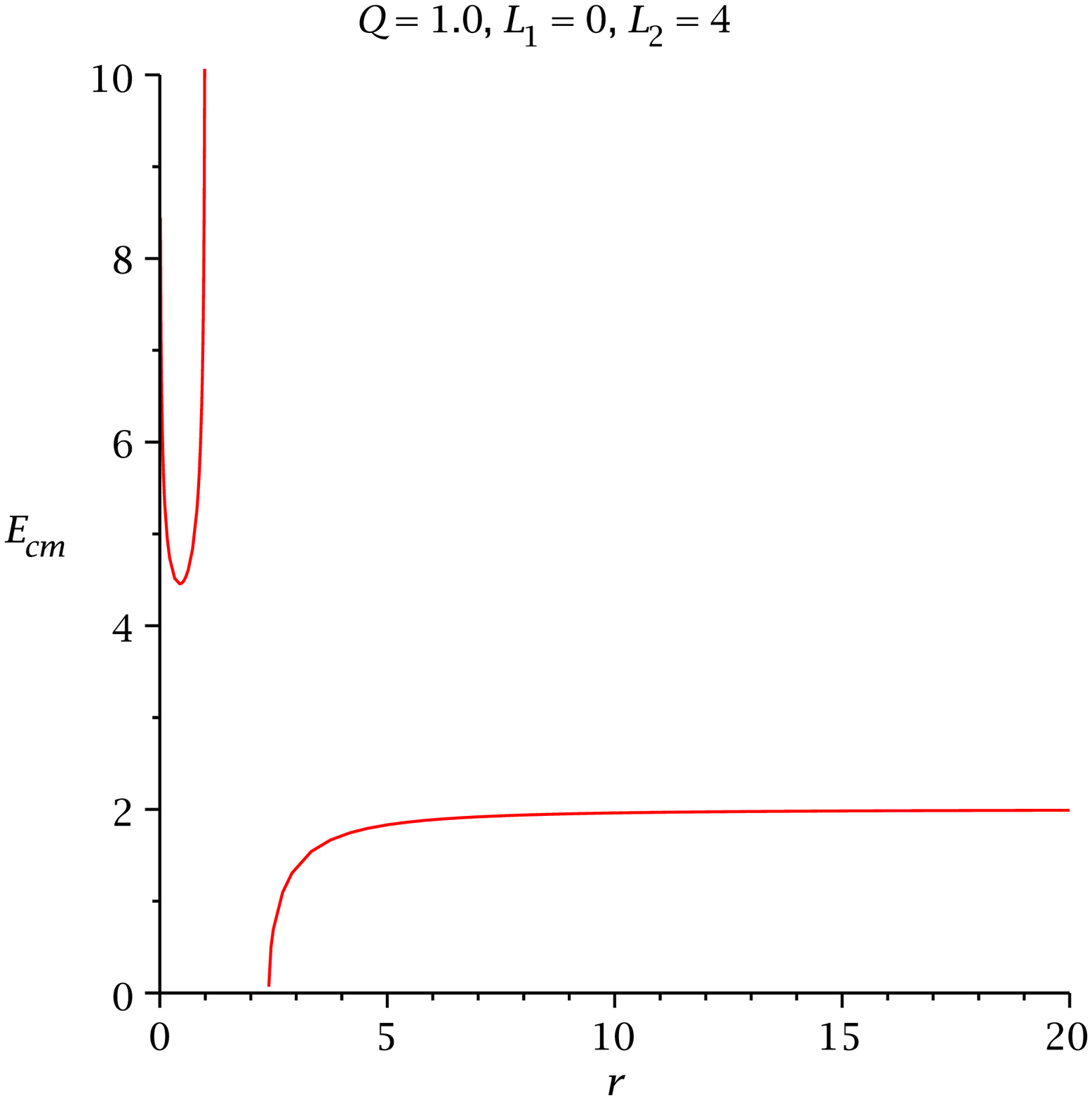}}\qquad
\subfigure[][]{\includegraphics[width=0.45\textwidth]{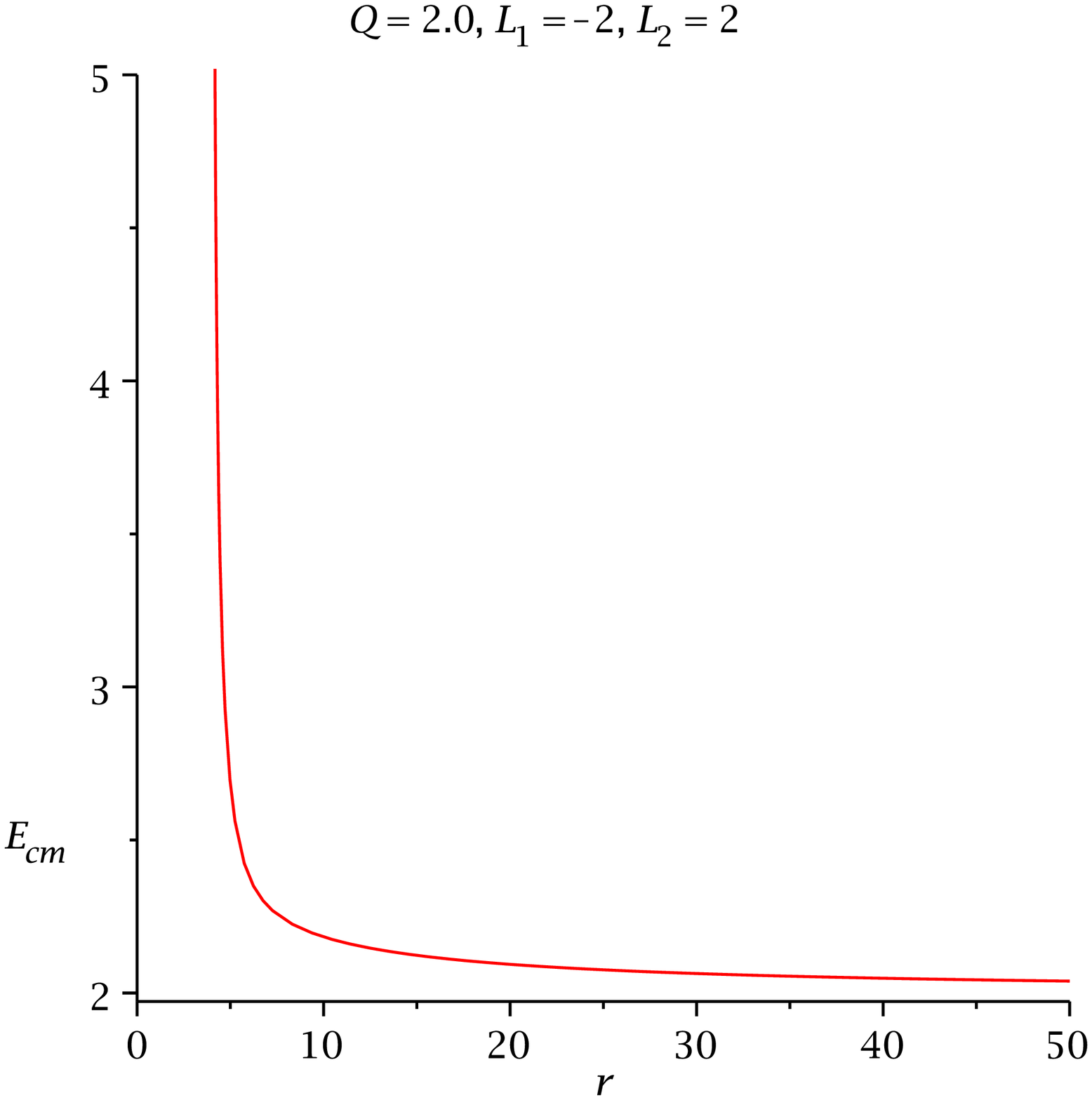}}
\end{center}
\caption{The figure shows the variation  of ${\cal E}_{cm}$  with $r$
for GMGHS BH.
 \label{gmgcase2}}
\end{figure}

\begin{figure}[t]
\begin{center}
\subfigure[][]{\includegraphics[width=0.45\textwidth]{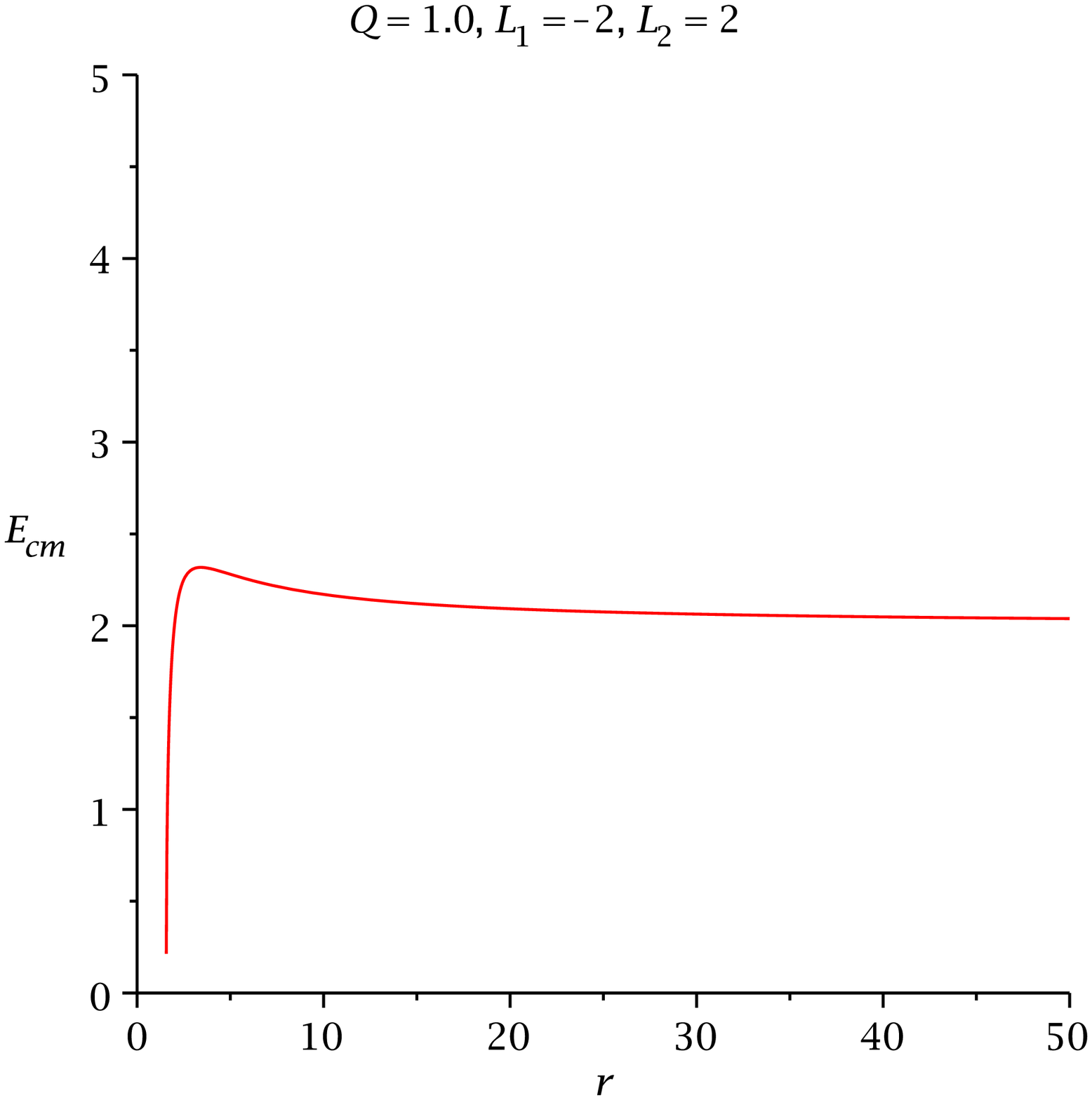}}\qquad
\subfigure[][]{\includegraphics[width=0.45\textwidth]{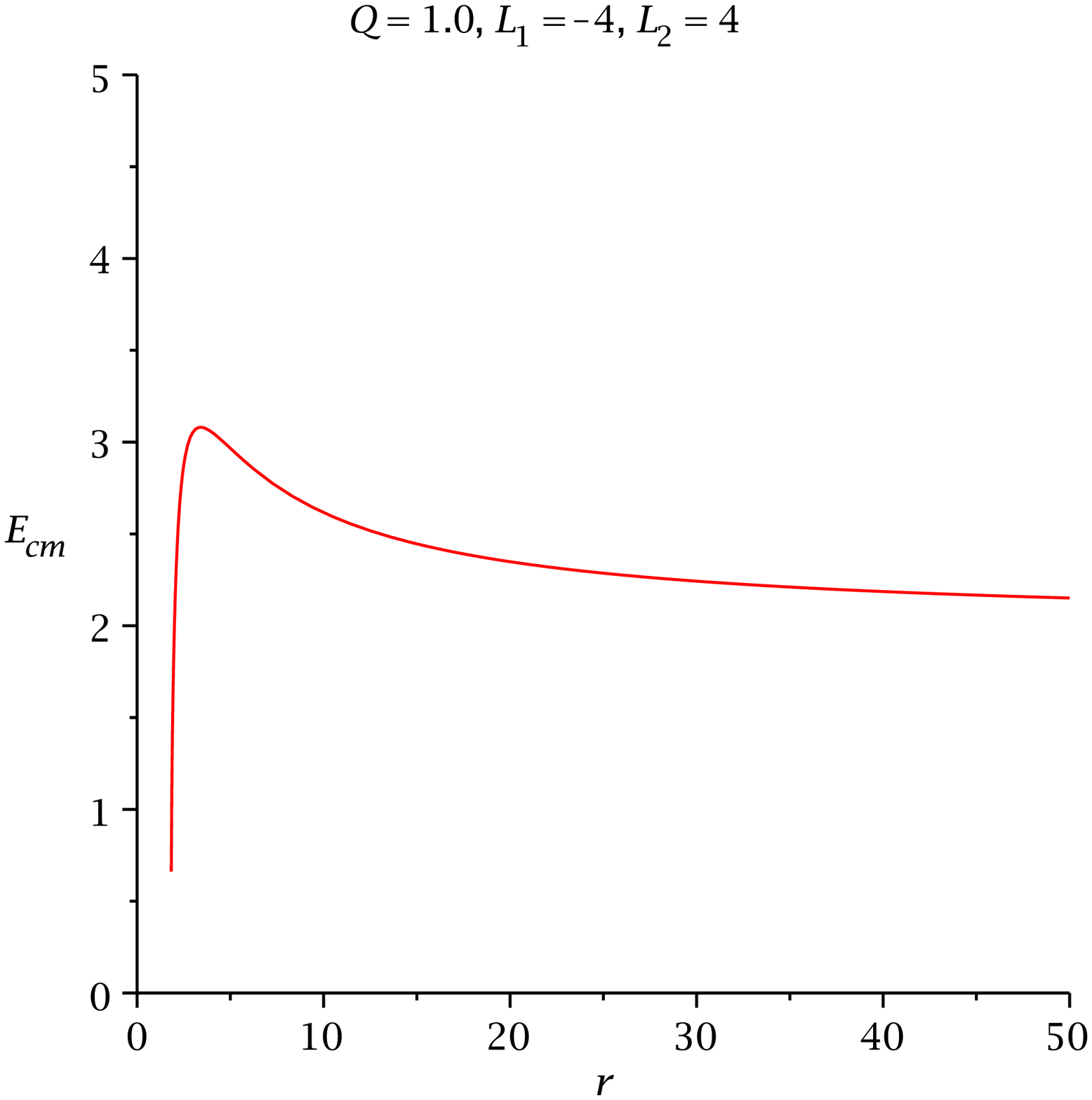}}
\end{center}
\caption{The figure shows the variation  of ${\cal E}_{cm}$  with $r$
for GMGHS BH.
 \label{gmgcase3}}
\end{figure}

Again, it is well known that for extremal GMGHS space-time \cite{ppgmghs}
the ISCO, circular photon orbit(CPO) and marginally bound circular orbit(MBCO) coincide
with the same radius i.e. $r_{ISCO}=r_{ph}=r_{mb}=2M$.
If we choose the different collision point say ISCO or CPO or MBCO, then the CM energy  will be unlimited for each collision point.
Now we have compared the ${\cal E}_{cm}$ for different collision point:
\begin{eqnarray}
{\cal E}_{cm}\mid_{r_{+}=2M}: {\cal E}_{cm}\mid_{r_{mb}=2M} : {\cal E}_{cm}\mid_{r_{ISCO}=2M}
&=& \infty : \infty : \infty
\end{eqnarray}
Since for this extreme BH, the collision point is at same location thus the
CM energy gives the diverging value at each collision point which is quite
different from extreme RN BH and Schwarzschild BH.

\section{Summary and Outlook:}

In this paper, we have performed the collision of two
neutral particles falling freely from rest at infinity in the
background  of the charged dilation BH. Firstly,
we have studied the complete geodesic structure of the charged
dilation BHs  both time-like case and null case respectively. ISCO, MBCO
and CPO of the said BHs  also have been computed.

Then we have discussed the extremal cases for both massive particles
and massless particles. Next, we have calculated the CM energy
in the center-of-mass frame for charged dilation BHs.
We have found that the ${\cal E}_{cm}$ is diverging for the extremal situation,
whereas ${\cal E}_{cm}$ is finite for the non-extremal situation. There are
two distinct outcome we have seen.  When  $n=1$, we recover the
value of ${\cal E}_{cm}$ for  RN BH and for extremal case, it gives the
diverging value. Again,  when  $n=1$, we retrieve  the value of
${\cal E}_{cm}$ for  GMGHS BH. Interestingly, for this extreme BH we
showed that in \cite{ppgmghs} the  ISCO, photon orbit and
marginally bound circular orbit were coincident with the event
horizon i.e $r_{ISCO}=r_{ph}=r_{mb}=r_{+}=2M$. Consequently, the
${\cal E}_{cm}$ at  $r \equiv r_{ISCO}=r_{ph}=r_{mb}=r_{hor}=2M$
gives the diverging value \cite{ppstring}. Which is completely different
from extreme RN BH, where, $r_{ISCO} \neq r_{ph} \neq r_{mb} \neq r_{+}=M$.

We also showed that for extreme RN BH the CM energy is
diverging at the extremal horizon and finite at the MBCO and ISCO. Their
ratio varies as ${\cal E}_{cm}\mid_{r_{+}=M}:  {\cal E}_{cm}\mid_{r_{mb}=\left(\frac{3+\sqrt{5}}{2}\right)M} : {\cal E}_{cm}\mid_{r_{ISCO}=4M} = \infty : 3.23 : 2.6$. While for Schwarzschild BH the ratio of CM energy is  ${\cal E}_{cm}\mid_{r_{+}=2M}:  {\cal E}_{cm}\mid_{r_{mb}=4M} : {\cal E}_{cm}\mid_{r_{ISCO}=6M} = \sqrt{5} : \sqrt{2} : \frac{\sqrt{13}}{3}$. We further showed that this ratio for GMGHS
BH varies as ${\cal E}_{cm}\mid_{r_{+}=2M}: {\cal E}_{cm}\mid_{r_{mb}=2M} : {\cal E}_{cm}\mid_{r_{ISCO}=2M} = \infty : \infty : \infty$.

\end{document}